\documentclass[prb,a4paper,reprint,superscriptaddress,preprintnumbers]{revtex4-2}

\usepackage{mymacros}
\definecolor{change}{rgb}{0.0, 0.0, 0.0} 
\newcommand{\change}[1]{{\textcolor{change}{#1}}}

\begin{document}

\title{Meson content of entanglement spectra after integrable and nonintegrable quantum quenches}

\author{Johannes Knaute} 
\email{johannes.knaute@mail.huji.ac.il}
\affiliation{Racah Institute of Physics, The Hebrew University of Jerusalem, Jerusalem 91904, Givat Ram, Israel}

\begin{abstract}
We use tensor network simulations to calculate the time evolution of the lower part of the entanglement spectrum and return rate functions after global quantum quenches in the Ising model. We consider ground state quenches towards mesonic parameter ranges with confined fermion pairs as nonperturbative bound states in a semiclassical regime and the relativistic E$_8$ theory. We find that in both cases only the dominant eigenvalue of the modular Hamiltonian fully encodes the meson content of the quantum many-body system or quantum field theory, giving rise to nearly identical entanglement oscillations in the entanglement entropy. When the initial state is prepared in the paramagnetic phase, the return rate density exhibits regular cusps at unequally spaced positions, signaling the appearance of dynamical quantum phase transitions, at which the entanglement spectrum remains gapped. Our analyses provide a deeper understanding on the role of quantum information quantities for the dynamics of emergent phenomena reminiscent of systems in high-energy physics.
\end{abstract}
\maketitle

\section{Introduction and motivation}

Quantum information concepts became increasingly relevant for the study of entanglement properties in strongly-coupled quantum many-body (QMB) systems and quantum field theories (QFTs) in and out of equilibrium \cite{Witten:2018zxz,Chen:2021lnq,Catterall:2022wjq}. While entanglement entropy is the most popular measure to quantify the amount of entanglement in pure states, to extract universal information, or to use it as an order parameter in (quantum) phase transitions (see e.g.\ the review \cite{Rangamani:2016dms}), the seminal paper \cite{Li:2008kda} introduced the more general \textit{entanglement spectrum}, \change{which allows to characterize the entanglement structure of a physical system in a pure state on an even deeper and more complete level}. 

Consider a pure state density operator $\rho$ and a spatial bipartition into a subsystem $A$ and its complement $B$. The \textit{modular} (or \textit{entanglement}) \textit{Hamiltonian} $\mathcal H_{mod}$ \cite{Haag:1992hx} is then defined from the reduced density matrix $\rho_A$ of the subsystem via
\begin{equation} \label{eq:H_mod}
    \rho_A = \Tr_B \rho \equiv \e^{-\mathcal H_{mod}} .
\end{equation}
The corresponding set of eigenvalues is denoted as the entanglement spectrum, from which the entanglement entropy and R\'enyi entropies can be calculated. While this concept was originally employed to detect topological order \cite{Li:2008kda,PhysRevB.84.205136}, it found enormous amount of attention across different fields in physics (see e.g.\ \cite{Dalmonte:2022rlo} for a review). In particular, it has been studied for lattice models \cite{Peschel_2009,PhysRevLett.108.227201,PhysRevLett.110.260403,Luitz_2014,Tonni:2017jom,Zhu:2018lua,Giudici:2018izb,DiGiulio:2019lpb,Mendes-Santos:2019tmf,Eisler:2020lyn,Rottoli:2022ego,Song:2022lxc} and fermionic systems \cite{Eisler_2017,Eisler_2018,ParisenToldin:2018uzz,Eisler:2019rnr,Fries:2019ozf}. Calculations of $\mathcal H_{mod}$ in QFTs, and especially conformal field theories (CFTs), are based on the Bisognano--Wichmann theorem \cite{Bisognano:1975ih,Bisognano:1976za}, which allowed to find some explicit forms \cite{Casini:2011kv,Wen:2016inm,Cardy:2016fqc,Wen:2018svb}. The modular Hamiltonian and its spectrum have also been studied using tensor networks \cite{Pollmann_2010,PhysRevB.83.245134,Schuch:2012bxy,Hsieh:2014jba,PhysRevLett.119.070401} and via holography in connection to further quantum information measures \cite{Casini:2011kv,Jafferis:2014lza,Jafferis:2015del}.

In this letter, we are interested in studying the impact of meson confinement on the dynamics of entanglement spectra after quantum quenches. Mesons are nonperturbative bound states, which appear in quantum chromodynamics (QCD) as flux tube confined quark-antiquark pairs that are important for the physics of the early universe after the big bang and heavy-ion collisions in nuclear accelerators \cite{Busza:2018rrf,Rothkopf:2019ipj,Berges:2020fwq}. The phenomenology of meson confinement, however, is not exclusive to QCD. Mesonic bound states exist also as confined fermion pairs (domain walls) in the spectrum of the quantum Ising model with longitudinal field \cite{McCoy:1978ta} or long-range interactions \cite{Liu:2018fza,Lerose_2019}. The seminal paper \cite{Kormos2017} initiated the study of their impact on the entanglement dynamics. Specifically, it was found that mesons give rise to \textit{entanglement oscillations}, i.e.\ an oscillating behavior of the entanglement entropy after quantum quenches, which bounds the overall entanglement growth if the quench is performed within the ferromagnetic phase and mesons are produced at rest. While analyses of quantum quenches towards critical regimes revealed that the entanglement spectrum carries universal information in form of the operator scaling dimensions of the underlying boundary CFT \cite{Cardy:2016fqc,Surace:2019mft,Robertson:2021wpp}, comparable studies in mesonic models have not yet been pursued. We fill this gap in this article using tensor network simulations \cite{Okunishi2022review,Banuls2022review} for both nonintegrable semiclassical and integrable relativistic regimes of the Ising model at early and intermediate time scales. 

We are particularly also interested in differences between quenches within the ferromagnetic phase versus crossings from the paramagnetic one. It hence becomes insightful to discuss our analyses in connection with \textit{dynamical quantum phase transitions} (DQPTs). These are non-equilibrium phase transitions, which occur in the time domain after quenches, showing up as nonanalyticities (cusps) in return rate functions. (For reviews on that topic see \cite{Heyl:2017blm,Heyl:2018jzi}.) Originally discovered through \textit{regular cusps} in \cite{PhysRevLett.110.135704} for quenches across the critical point of the transverse field Ising model, it was realized that DQPTs exist also for phase crossings in the longitudinal field \cite{PhysRevB.87.195104} and long-range Ising model \cite{PhysRevLett.120.130601,PhysRevB.96.134427}, i.e.\ in models where mesons can exist. Their appearance was experimentally confirmed in \cite{PhysRevLett.119.080501,Zhang_2017}. Moreover, it was shown that \textit{anomalous} DQPTs can even exist for quenches within the ferromagnetic phase \cite{PhysRevB.96.134427,PhysRevE.96.062118,Halimeh_2020,hashizume2020dynamical,Defenu:2019dkd,halimeh2021dynamical}. Connections between DQPTs and the dynamics of the entanglement spectrum have been pioneered in \cite{PhysRevB.89.104303,Torlai_2014,DeNicola:2020ddc,PhysRevA.103.012204}. While the necessity of meson states for anomalous DQPTs has been explored in \cite{Halimeh_2020,Defenu:2019dkd}, their explicit role in the entanglement spectrum, however, has not yet been addressed.

Our analyses are also strongly motivated by significant advances of quantum simulation technologies for the study of fundamental physics problems \cite{Alexeev:2020xrq,Bauer:2022hpo,Fraxanet:2022wgf,Daley:2022eja}. Recently, the impact of confinement and mesons on quantum correlations, entanglement dynamics and related properties has been studied experimentally \cite{Tan:2019kya,Mildenberger:2022jqr} and theoretically \cite{Verdel:2019chj,Surace:2020ycc,Karpov:2020pqe,Rigobello:2021fxw,Knaute:2021xna,Halimeh:2022pkw,Banuls:2022iwk,Kebric:2022pml,Werner:2022kei}. On the other hand, not only DQPTs became accessible in quantum simulations \cite{PhysRevLett.119.080501,Zhang_2017}, but also the spectrum of the modular Hamiltonian via entanglement tomography \cite{Dalmonte:2017bzm,Kokail:2020opl,Kokail:2021ayb}. It therefore is a very timely problem to address the impact of meson confinement also in the latter context.

\section{Model}

The one dimensional nearest-neighbor quantum Ising model is defined by the Hamiltonian
\begin{equation}  \label{eq:H_NN}
H = - J \,\left ( \sum_{j=1}^{N-1} \sigma^z_j \sigma^z_{j+1} + h \sum_{j=1}^N  \sigma^x_j + g \sum_{j=1}^N \sigma^z_j \right) ,
\end{equation}
where $\sigma^\alpha_j$ ($\alpha=\{x,z\}$) are Pauli matrices at lattice position $j$ within an open chain of $N$ sites. The unit $J \equiv 1$ sets the overall lattice energy scale, and the transverse and longitudinal field perturbations w.r.t.\ the first interaction term are quantified by the parameters $h$ and $g$, respectively. The transverse model ($g=0$) exhibits a quantum critical point at $J\!=\!h\!=\!1$, at which a quantum phase transition from a disordered paramagnetic phase ($h>1$) towards an ordered ferromagnetic phase ($h<1$) occurs \cite{sachdev_2011}. 

In the thermodynamics limit ($N \to \infty$), there exists a scaling limit, in which the infrared regime is described by a Majorana fermion QFT, given by the Hamiltonian~\cite{Rakovszky:2016ugs}
\small
\begin{equation} \label{eq:H_IsingQFT}
H_\text{IR} = \int_{-\infty}^\infty \mathrm dx \,  \left\{ \frac{i}{4\pi} \left( \psi\partial_x\psi - \bar\psi\partial_x\bar\psi \right) - \frac{i M_h}{2 \pi}\bar\psi\psi + {\cal C} M_{g}^{15/8} \, \sigma \right\} .
\end{equation}
\normalsize
Here, $M_{h}\!\equiv\!2 J |1-h|$ is the free fermion mass, $M_{g} \equiv \, {\cal D} J \, |g|^{8/15}$ is a longitudinal mass scale, and ${\cal C} \approx 0.062, {\cal D} \approx 5.416$ are numerical constants~\cite{Rakovszky:2016ugs, Hodsagi:2018sul}. \change{The spin field $\sigma$ is the continuous generalization of $\sigma_j^z$.}

At criticality, i.e.\ for $M_h = M_g = 0$, the Hamiltonian \eqref{eq:H_IsingQFT} describes the Ising conformal field theory (CFT) of central charge $c=1/2$, which possesses two scalar primary operators, $\epsilon=i\bar\psi\psi$ and $\sigma$ with scaling dimensions $\Delta_\epsilon=1$ and $\Delta_\sigma=1/8$. Transverse perturbations of the Ising CFT ($M_h>0$, $M_g=0$) result in an integrable massive free fermion regime. Longitudinal perturbations confine domain walls as elementary excitations in the ferromagnetic phase into nonperturbative meson bound states~\cite{McCoy:1978ta}. In particular, pure longitudinal perturbations ($M_h=0$, $M_g>0$) give rise to the integrable and interacting E$_8$ QFT \cite{Zamolodchikov:1989fp}, whose 8 stable meson masses $M_n$ are analytically known as ratios to the lightest mass $M_1 \equiv M_g$. Combined transverse and longitudinal perturbations ($M_h>0$, $M_g>0$) result in a nonintegrable interacting QFT with both stable and unstable mesonic bound states~\cite{Delfino:1996xp,Delfino2005Jul,Fonseca:2006au,Zamolodchikov:2013ama}.

\section{Setup}

In the present letter, we study real-time properties of entanglement spectra and return rate functions after global quantum quenches in both the integrable E$_8$ QFT as well as in the nonintegrable meson regime. For this purpose, we employ well established ab initio tensor network simulations, which directly give access to the quantities of interest in the thermodynamic limit of a translational invariant spin chain. In particular, based on the matrix product state (MPS) ansatz \cite{Verstraete08,SCHOLLWOCK201196,Cirac2009,Orus:2013kga}, we use the infinite time-evolving block decimation (iTEBD) algorithm \cite{Vidal:2006ofj} to construct a MPS approximation to (gapped) ground states $\ket{\psi_0} = \lim_{\beta\to\infty}\e^{-\beta H_0}$ w.r.t.\ an Ising model Hamiltonian $H_0$ of the form \eqref{eq:H_NN} via imaginary time evolution. We then use the same iTEBD algorithm to calculate its real-time evolution $\ket{\psi(t)} = \e^{-itH_1}\ket{\psi_0}$ under a different Hamiltonian $H_1$. In nontrivial cases, the state $\ket{\psi_0}$ is not an eigenstate of $H_1$, such that this quench protocol drives the QMB system instantaneously out-of-equilibrium (at time $t=0$) and causes the emergent phenomena.

\begin{figure}[t]
    \centering
   \includegraphics[width=0.65\columnwidth]{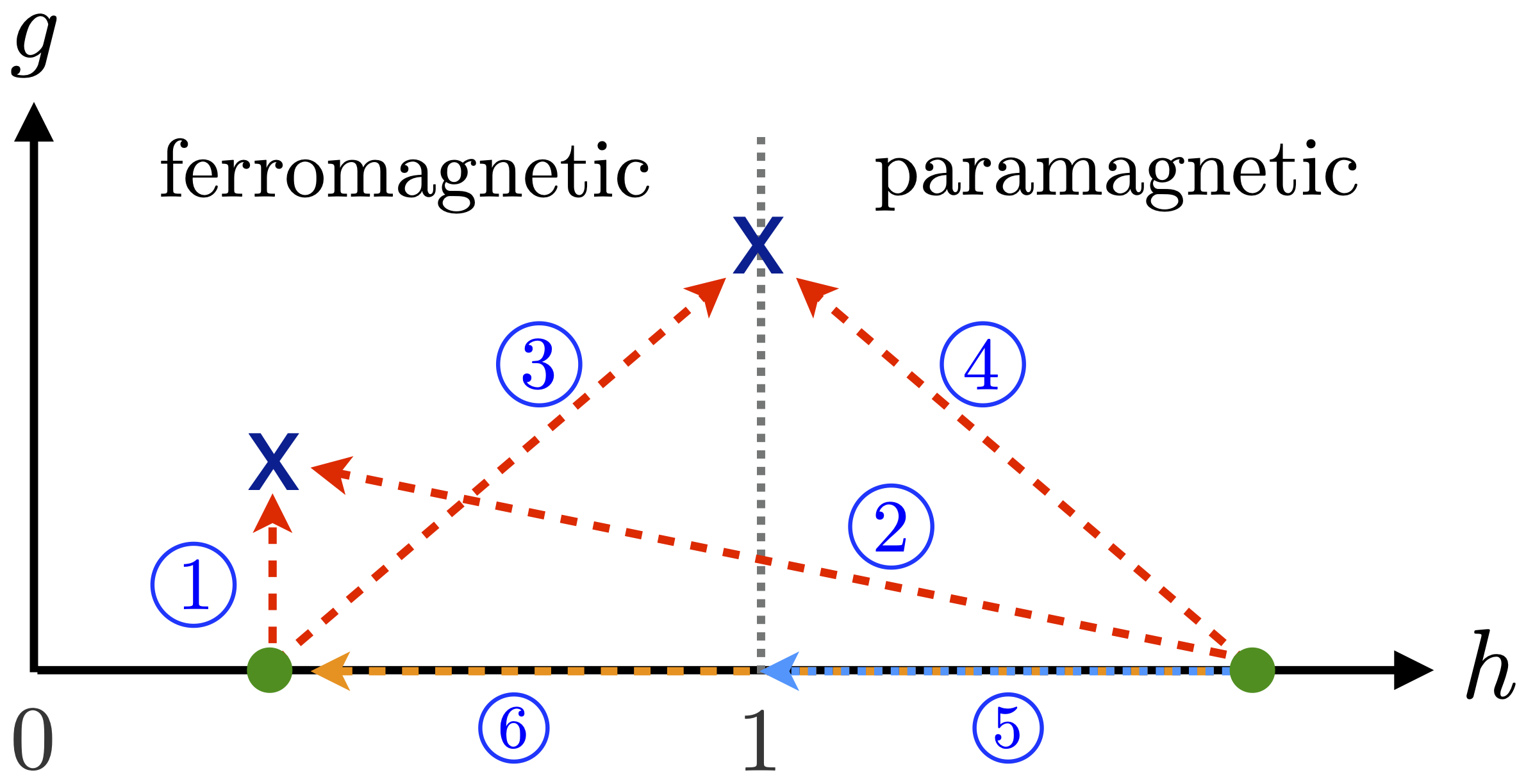} 
    \caption{Overview of the considered quench protocols in the transverse ($h$) vs.\ longitudinal ($g$) field plane. Ground states are prepared in the ferromagnetic and paramagnetic phase of the purely transverse field Ising model (indicated by green dots) and quenched towards a nonintegrable semiclassical meson regime [types (1,2)] and the integrable E$_8$ QFT regime (indicated by the grey dotted line) [types (3,4)].}
    \label{fig:QuenchTypes}
\end{figure}

We consider the specific quench protocols illustrated in Fig.\,\ref{fig:QuenchTypes}. We choose two distinct pre-quench parameter points in the free fermion ferromagnetic and paramagnetic phase (shown as green dots) for the parameters $h=0.25$ and $h=1.75$, respectively. Protocols \circled{1} and \circled{2} quench towards a nonintegrable meson regime for which we exemplarily choose $\{h=0.25,g=0.1\}$ (indicated by the left cross). This quench point is far away from criticality, i.e.\ a QFT description is not amenable but instead a semiclassical approximation based on the Bohr--Sommerfeld quantization condition can be used to determine four meson states and their masses (see \cite{Kormos2017} for detailed discussions). Protocols \circled{3} and \circled{4}, on the other hand, quench to the integrable E$_8$ QFT regime from the different pre-quench phases. The post-quench parameter point is given for $\{h=1,g=0.48\}$.\,\footnote{The values are chosen such that both final quench points have an identical mass gap, given by the first meson mass $M_1/J \approx 3.66$.} \change{In App.\,\ref{app:critical} \footnote{See the supplemental material below for several appendices containing further background material.} we contrast the resulting properties to a non-mesonic case, realized through quenches from the paramagnetic phase to the critical point (protocol \circled{5}, CFT results are available) and towards the ferromagnetic phase in the free fermion regime (protocol \circled{6}, regular DQPTs occur).}

We analyze real-time entanglement properties of the state $\ket{\psi(t)}$ for a semi-infinite bipartition of the Ising chain, \change{realized through a cut in between two repeating tensors of the translational invariant chain, which defines subsystem A as all the infinitely many sites to the left of the cut, and the complement B as all sites to the right.} A Schmidt decomposition across this cut takes the form $\ket{\psi(t)} = \sum_r \sqrt{\lambda_r} \ket{\psi^A_r}\otimes\ket{\psi^B_r}$, where the Schmidt values $\lambda_0 \ge \lambda_1 \ge \lambda_2 \ge \ldots$ are directly related to the eigenvalues $\xi_r$ of the entanglement spectrum via $\lambda_r \equiv \e^{-\xi_r}$. A relevant quantity of interest are the entanglement gaps $g_r \equiv \ln\lambda_0 - \ln\lambda_r = \xi_r - \xi_0$. The entanglement entropy is given by $S_1(\rho_A) = -\Tr_A[\rho_A \ln\rho_A] = -\sum_r \lambda_r \ln\lambda_r$ \change{as the von-Neumann entropy of the reduced density matrix}, and the 2-R\'enyi entropy follows as $S_2(\rho_A) = -\ln\Tr_A\rho_A^2 = -\ln\sum_r\lambda_r^2$.

The central quantity to identify DQPTs is the Loschmidt amplitude $G(t) \equiv \braket{\psi(0)|\psi(t)} = \braket{\psi(0)|\e^{-itH_1}|\psi(0)}$, from which the \textit{return rate density} is defined as 
\begin{equation}
    r_1(t) = -\lim_{N\to\infty} \frac{1}{N} \ln\vert G(t)\vert^2 .
\end{equation}
The latter can be interpreted as an analogue of the free energy density in equilibrium, such that nonanalyticities in $r_1(t)$ indicate the appearance of DQPTs as dynamical analogues of equilibrium phase transitions \cite{PhysRevLett.110.135704,Heyl:2017blm,Heyl:2018jzi}. As discussed in \cite{PhysRevE.96.062118}, this definition can be generalized to the \textit{rate functions}
\begin{equation}
    r_i(t) = -2 \ln\vert \epsilon_i(t) \vert .
\end{equation}
\change{Here, $\epsilon_i$ are the eigenvalues (in decreasing order) of the mixed MPS transfer matrix $\mathcal E(t) \equiv \Tr_{\text{phys}}[\bar C(0) \otimes C(t)]$ between two MPS tensors $C$ of $\ket{\psi_0}$ and $\ket{\psi(t)}$, where a trace over physical indices is taken.} Cusps or kinks in $r_1(t)$ correspond to level crossings between $r_1$ and $r_{i>1}$.

\section{Quenches to nonintegrable semiclassical meson regimes}

\begin{figure}[t]
    \centering
   \includegraphics[width=0.49\columnwidth]{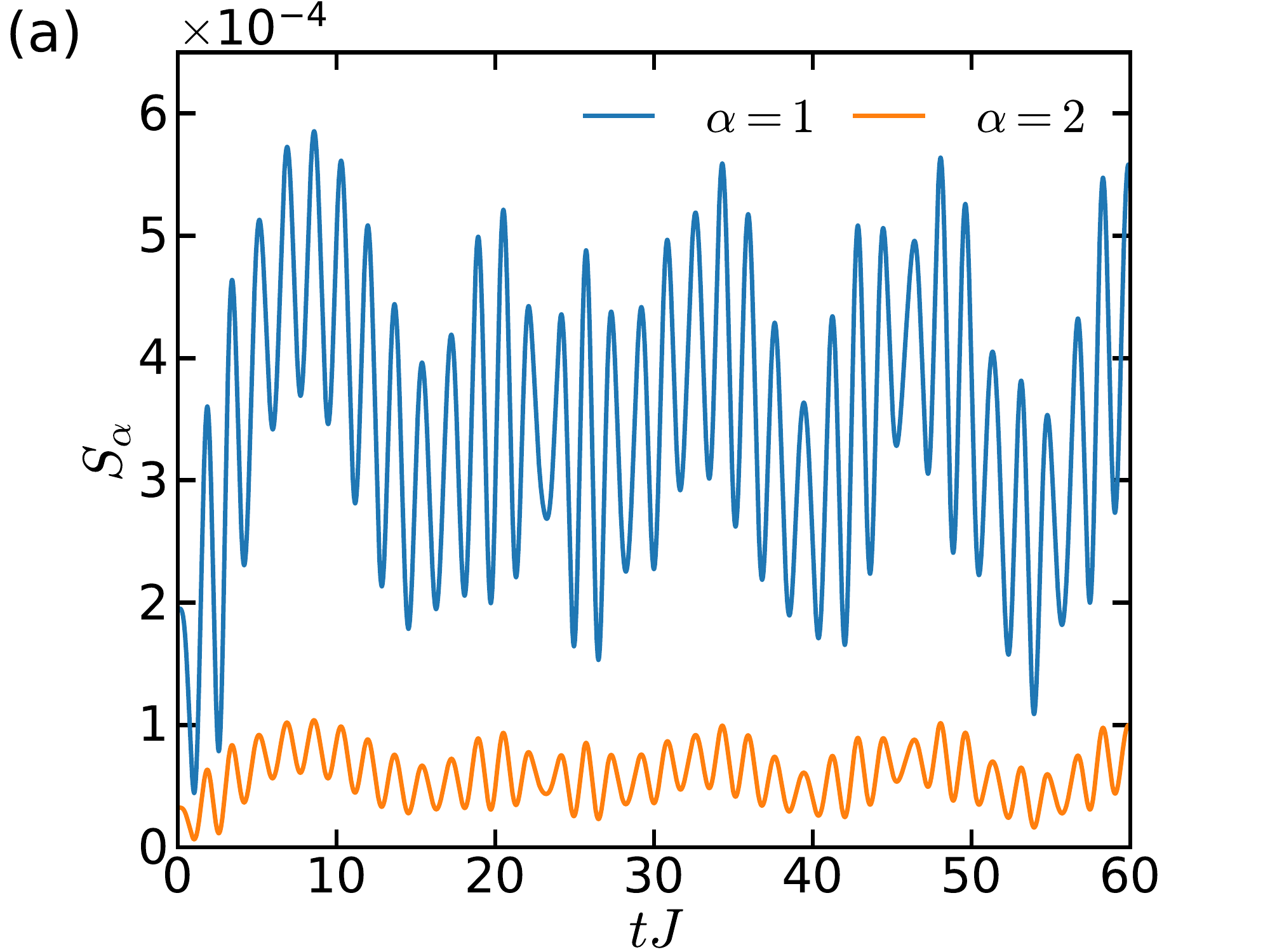} 
   \includegraphics[width=0.49\columnwidth]{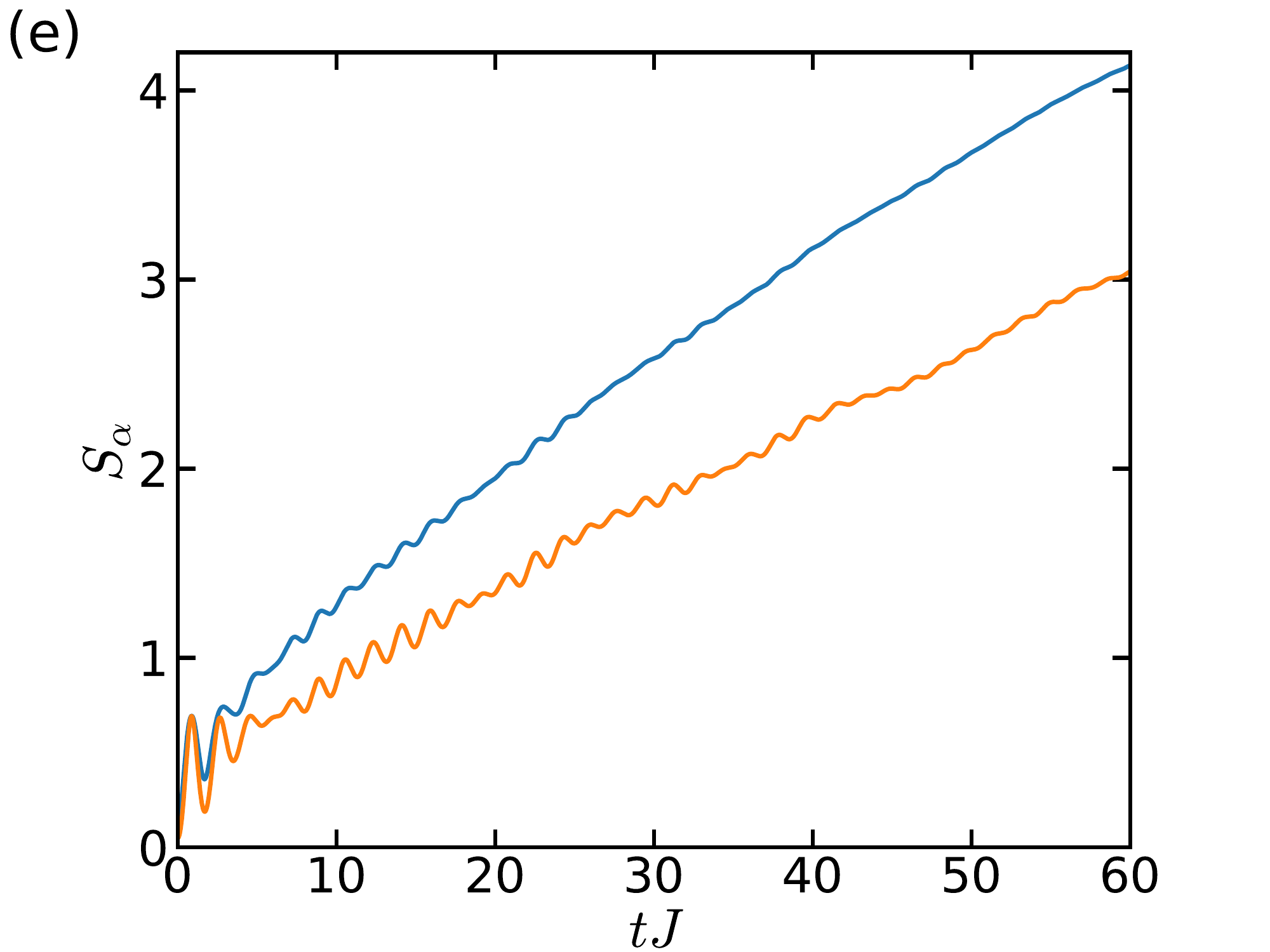} 
   \includegraphics[width=0.49\columnwidth]{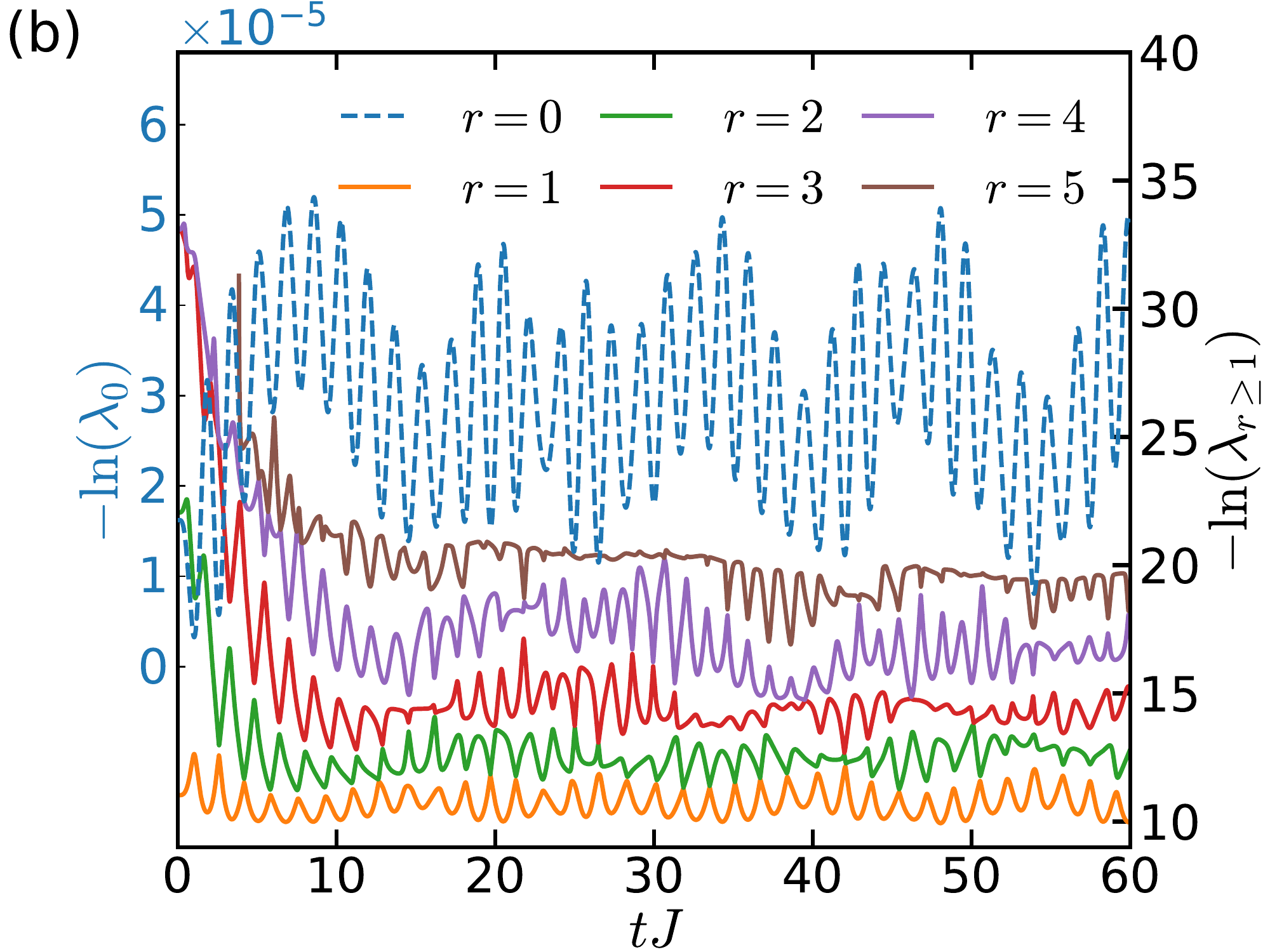} 
   \includegraphics[width=0.49\columnwidth]{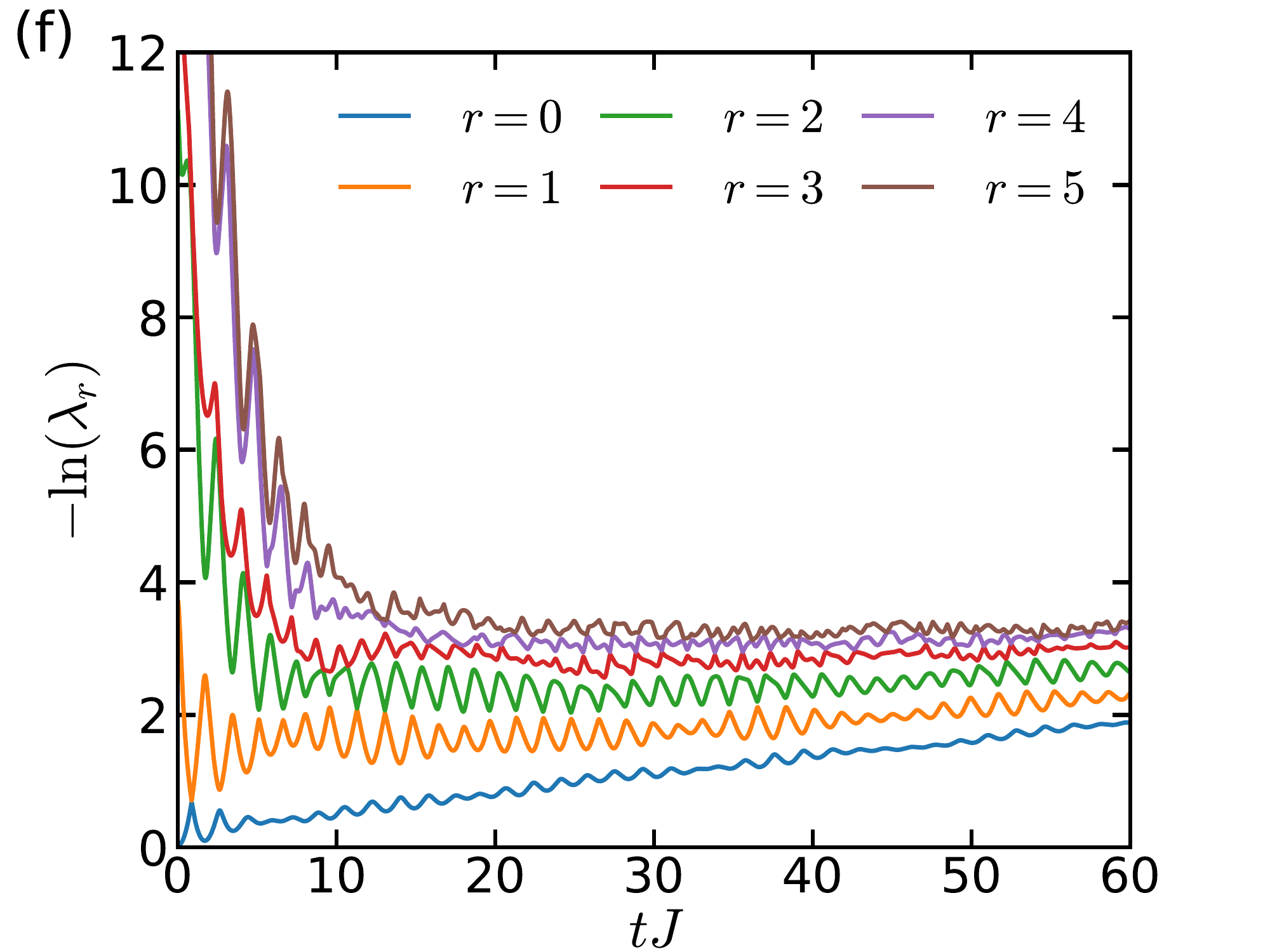} 
   \includegraphics[width=0.49\columnwidth]{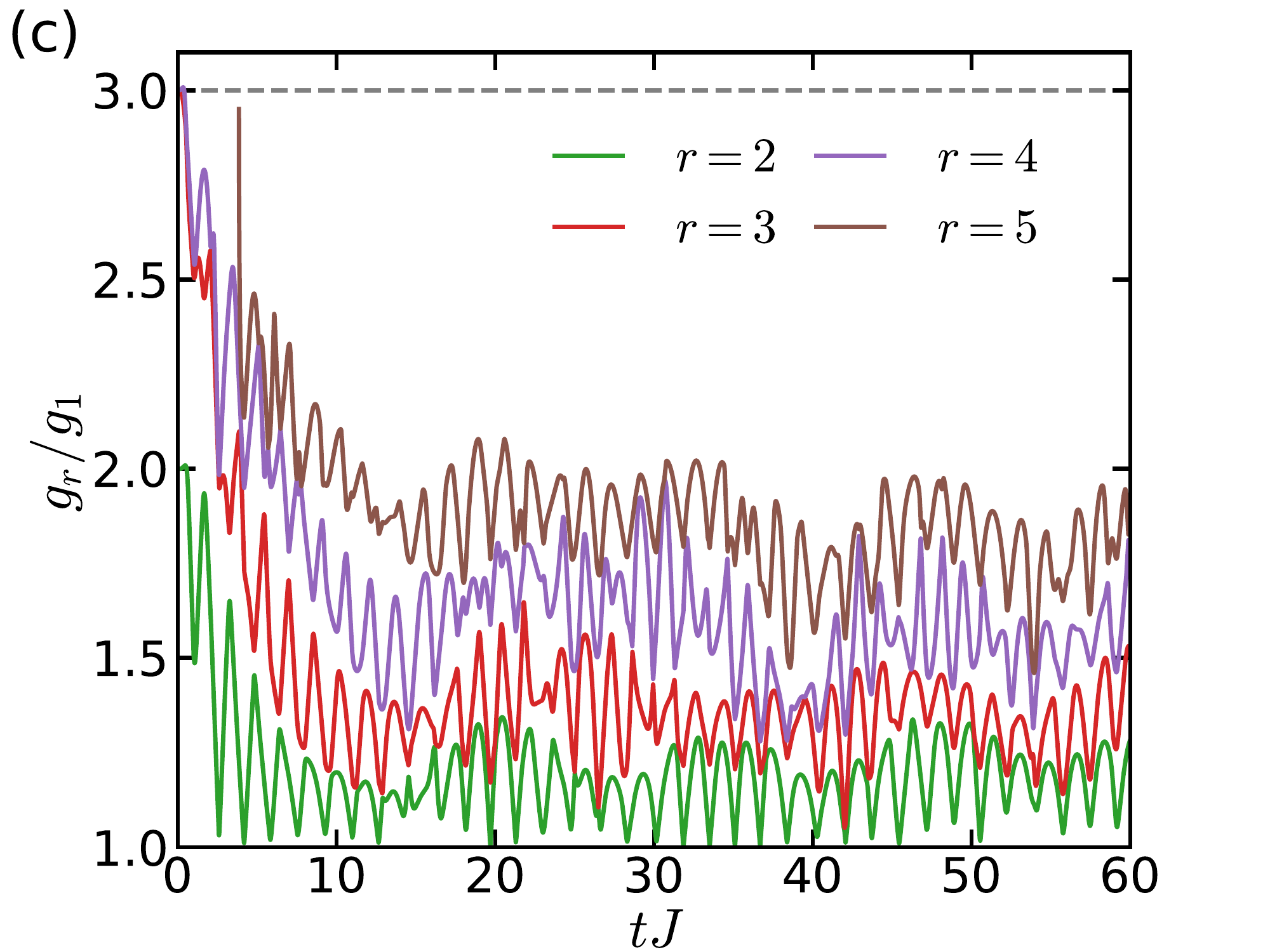} 
   \includegraphics[width=0.49\columnwidth]{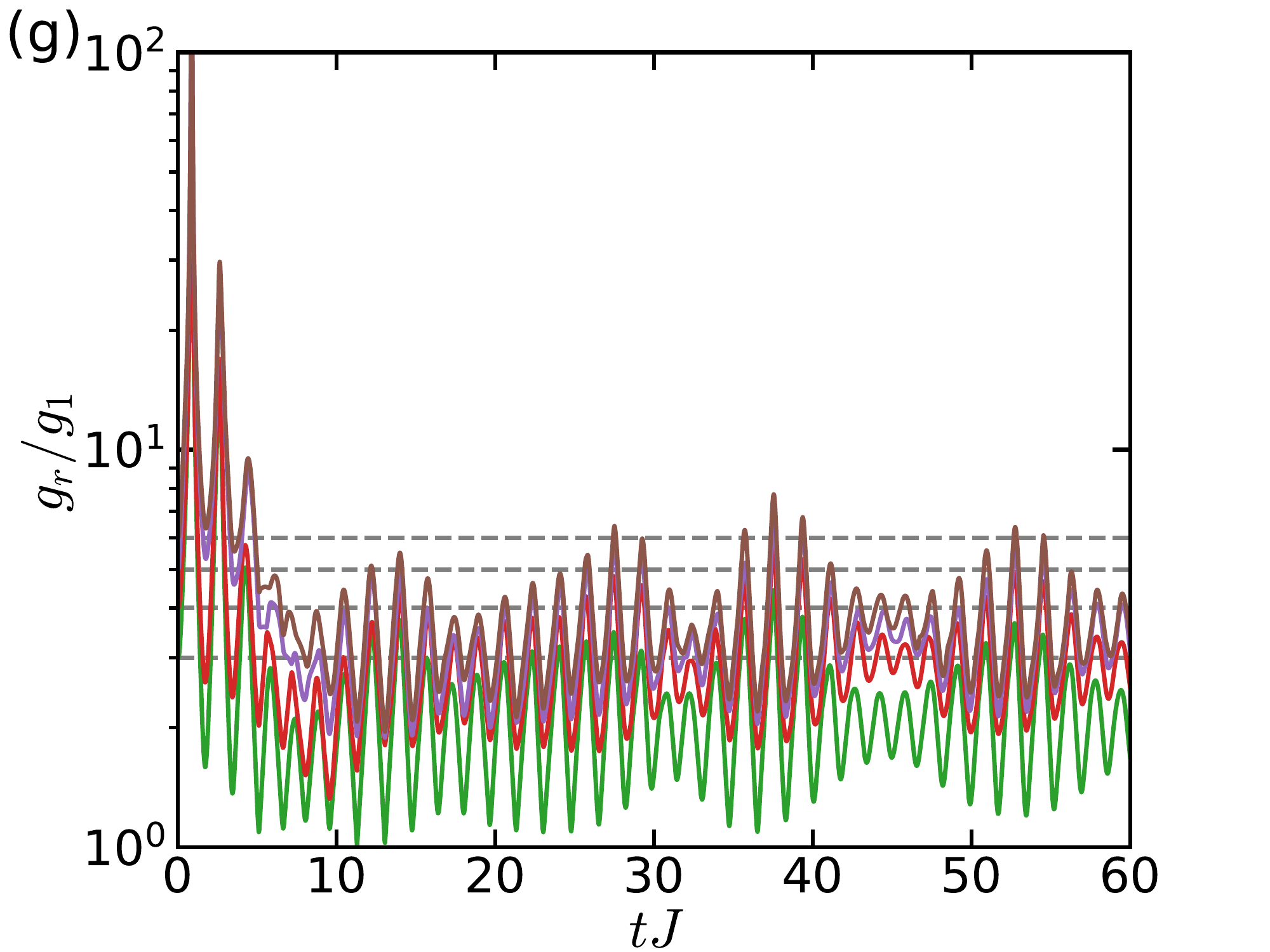} 
   \begin{minipage}{0.49\columnwidth}
   \includegraphics[width=\textwidth]{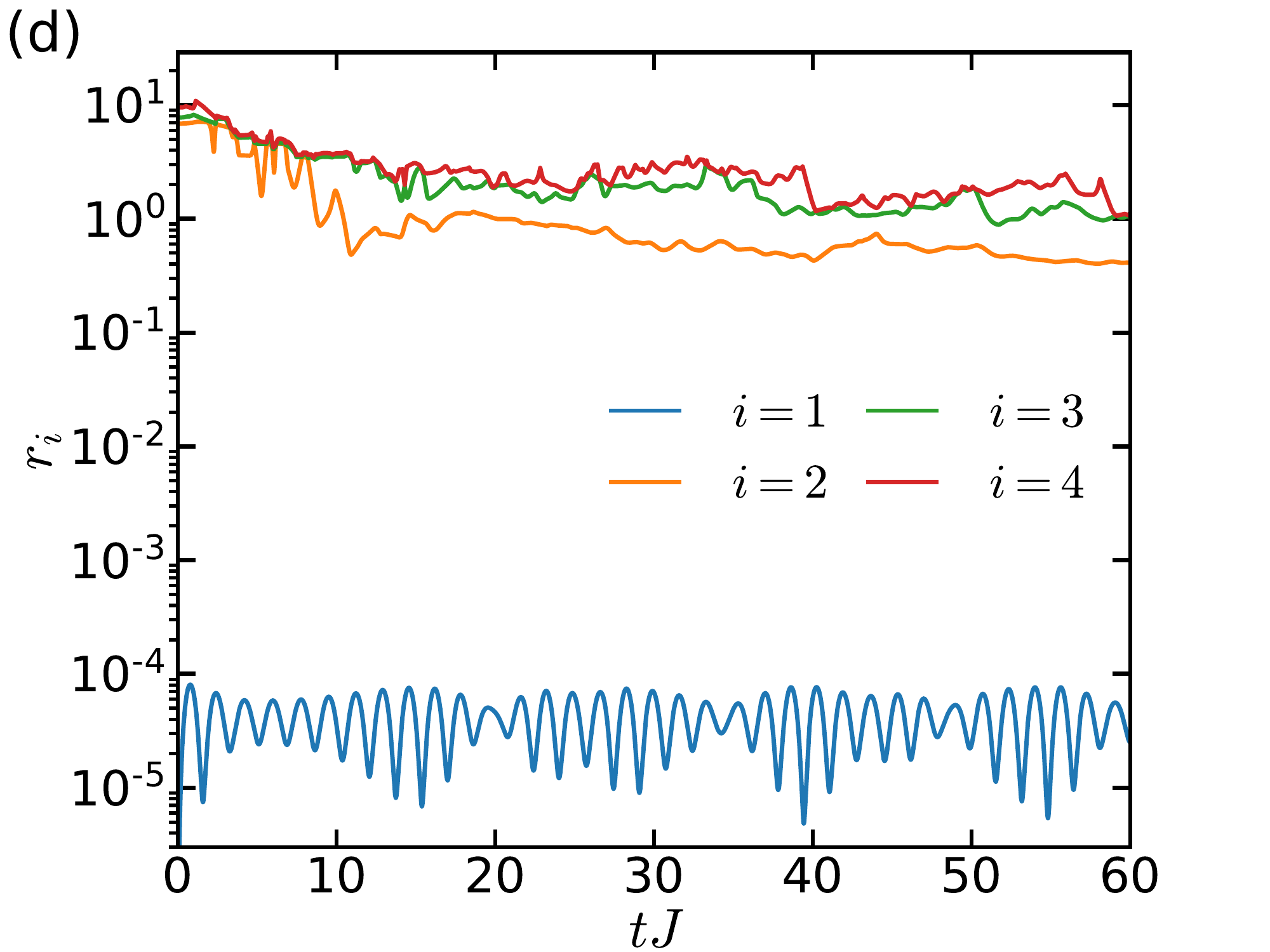} \\ 
   \circled{1}
   \end{minipage}
   \begin{minipage}{0.49\columnwidth}
   \includegraphics[width=\textwidth]{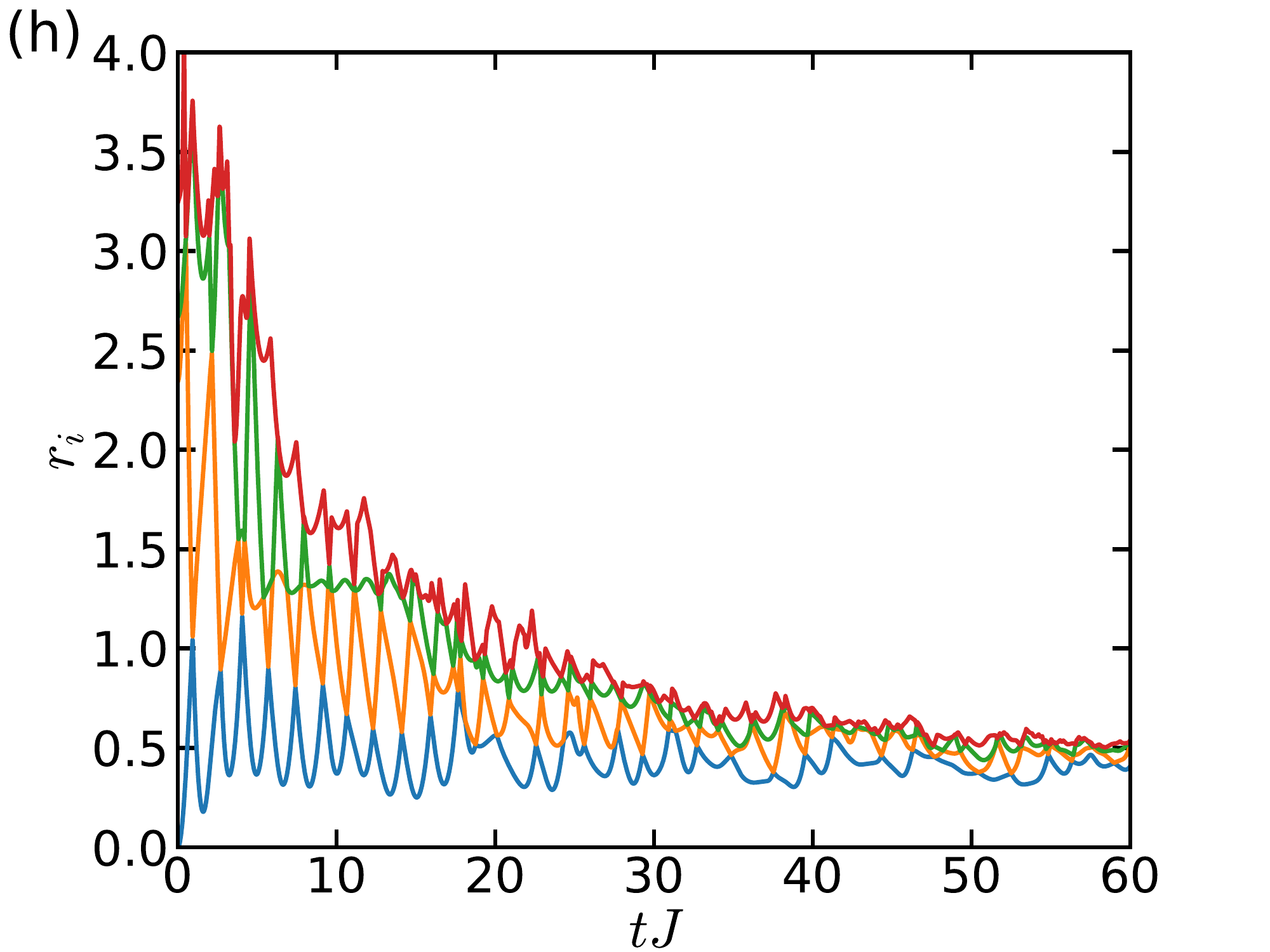} \\ 
   \circled{2}
   \end{minipage}
    \caption{Time dependence of physical quantities in quench protocols (1) (a-d) and (2) (e-h) to a semiclassical meson regime. From top to bottom: entanglement entropy $S_1$ and 2-R\'enyi entropy $S_2$ (a,e), eigenvalues $\lambda_r$ of the entanglement spectrum (b,f), entanglement gap ratios $g_{r\ge2}/g_1$ (c,g), return rate functions $r_i$ (d,h). See text for detailed discussions.}
    \label{fig:spectra_meson}
\end{figure}
\begin{figure}[t]
    \centering
   \includegraphics[width=0.49\columnwidth]{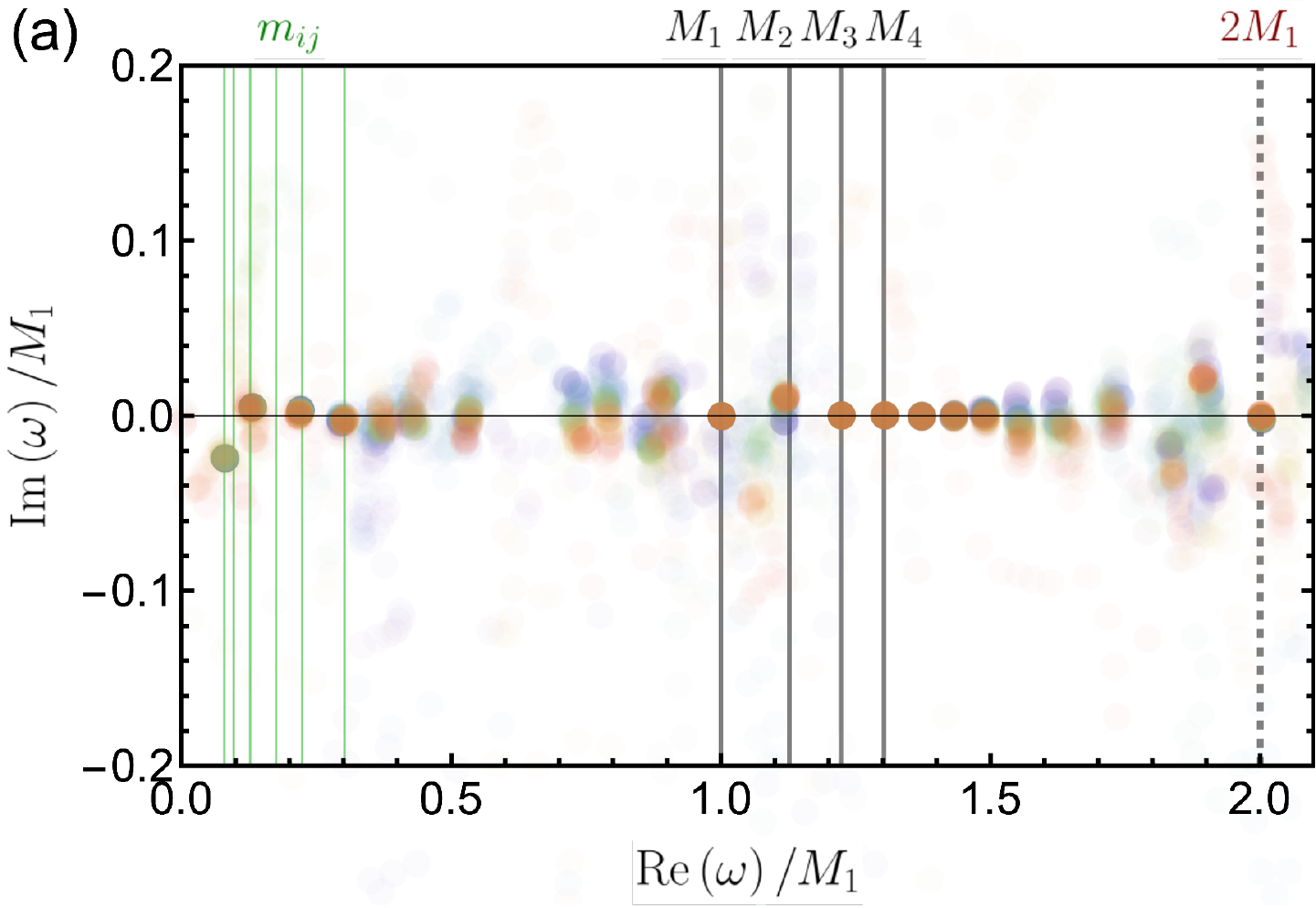} 
   \includegraphics[width=0.49\columnwidth]{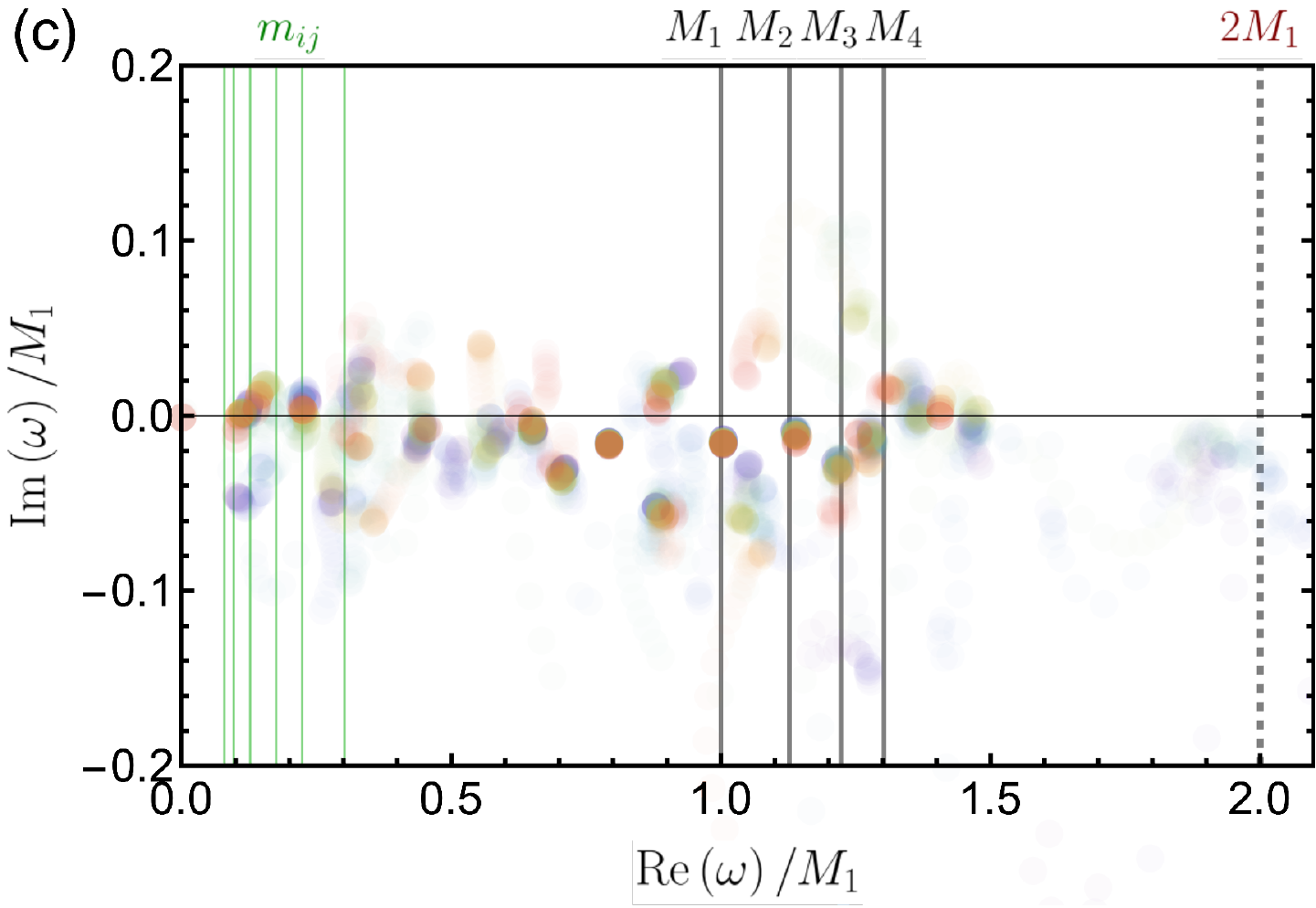} 
   \begin{minipage}{0.49\columnwidth}
   \includegraphics[width=\textwidth]{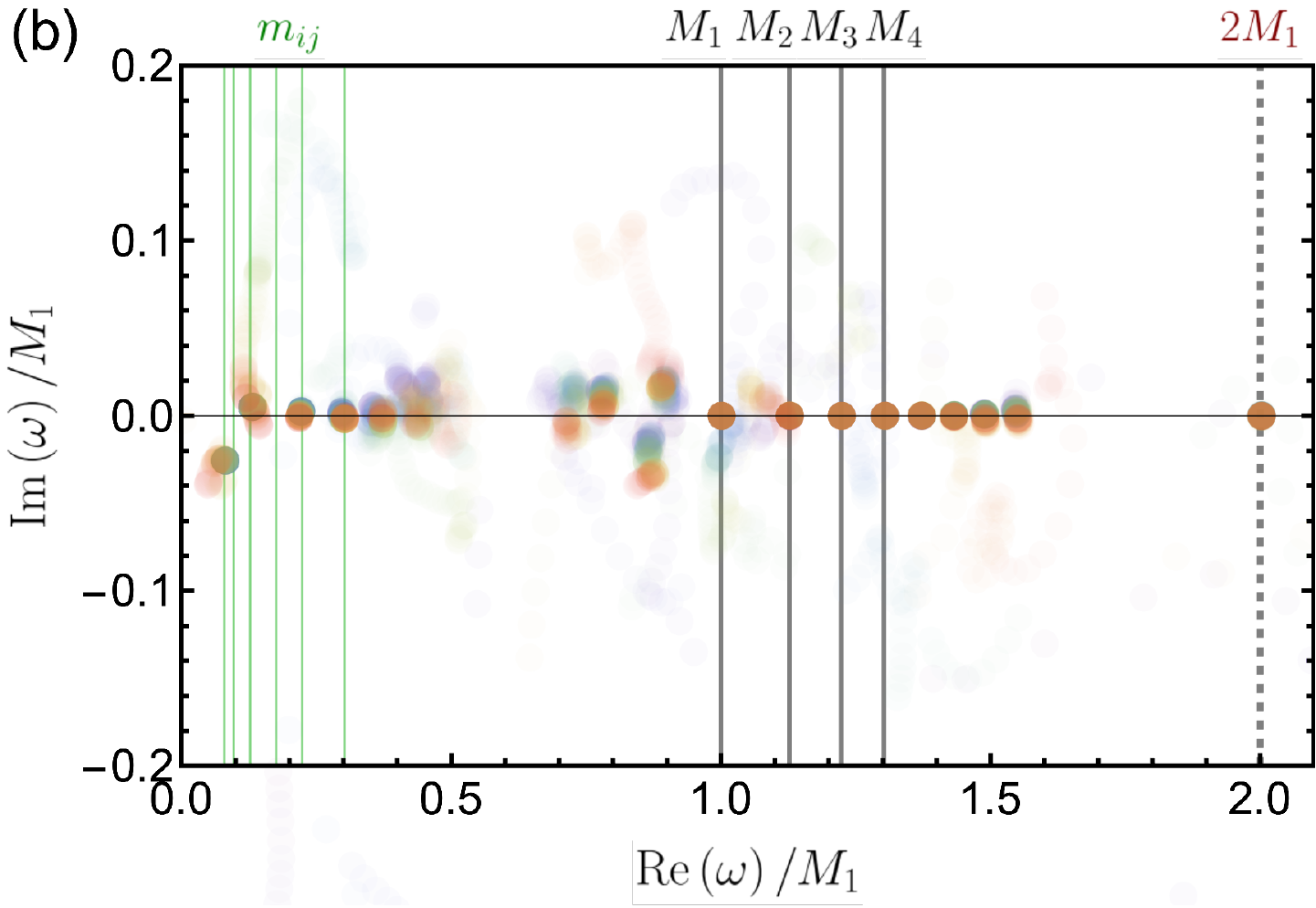}\\ 
   \circled{1}
   \end{minipage}
   \begin{minipage}{0.49\columnwidth}
   \includegraphics[width=\textwidth]{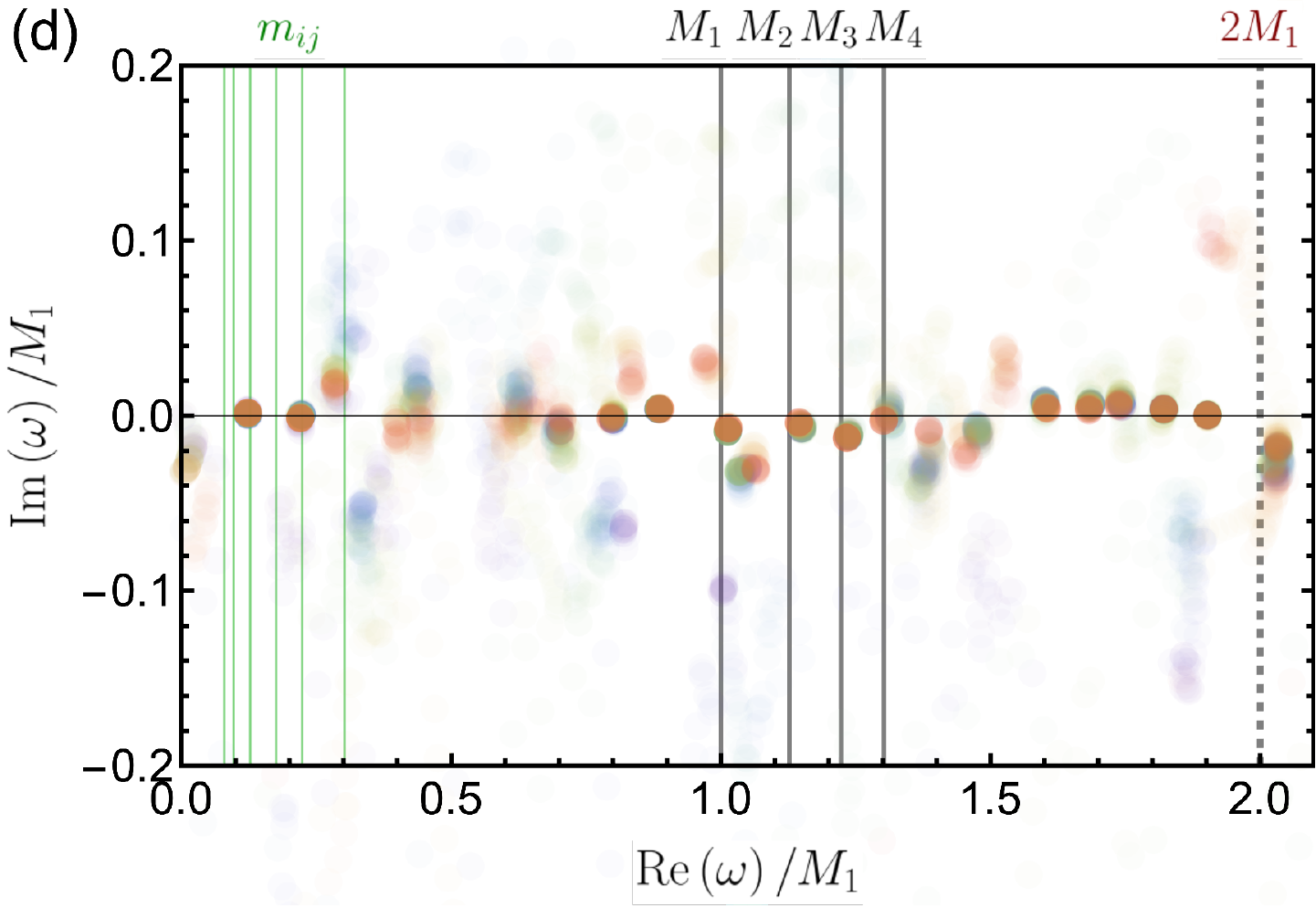}\\ 
   \circled{2}
   \end{minipage}
    \caption{Results of the Prony signal analysis of $S_1$ (a,c) and $\xi_0=-\ln(\lambda_0)$ (b,d) under quench type (1) (left column) and (2) (right column). Grey vertical lines indicate meson masses $M_i$ obtained from a semiclassical approximation, green vertical lines show all possible mass differences $m_{ij} \equiv |M_i-M_j|$ between them: \change{$m_{34}$, $m_{23}$, $m_{12}$, $m_{24}$, $m_{13}$, $m_{14}$ (ascending)}. The results demonstrate that the meson content of entanglement oscillations is fully encoded in the dominant eigenvalue of the modular Hamiltonian.}
    \label{fig:Prony_type12}
\end{figure}

Fig.\,\ref{fig:spectra_meson} shows the simulation results for quenches from the ferromagnetic (type \circled{1}, left column) and paramagnetic phase (type \circled{2}, right column) into the nonintegrable semiclassical meson regime. The time evolution of $S_1$ and $S_2$ for quench \circled{1} [panel (a)] within the ferromagnetic phase exhibits a bounded oscillatory behavior, representing the known entanglement oscillations induced through meson confinement \cite{Kormos2017}.\,\footnote{We have chosen the parameters in quench protocol (1) identical to one of the cases in \cite{Kormos2017} and hence reproduce the form of $S_1$ here.} On the other side, $S_1$ and $S_2$ show a very large entanglement growth under quench \circled{2} [panel (e)], which are superimposed with oscillations. The latter are, in contrast, unbounded in the available simulation times.\,\footnote{We refer to App.\,\ref{app:iTEBD} for some details of the iTEBD simulations.} 

Panels (b) and (f) show the first eigenvalues $\xi_0,\ldots,\xi_5$ of the corresponding entanglement spectra. One can observe that the dominant eigenvalue $\xi_0 = -\ln(\lambda_0)$ in quench type \circled{1}, shown as the blue dashed curve in panel (b), oscillates on a much smaller magnitude (w.r.t.\ the left axis) than the remaining eigenvalues, i.e.\ the entanglement spectrum is largely gapped. The shape of $\xi_0$ follows nearly identically the time evolution of $S_1$ and hence seems to encode the entanglement oscillations (cf.\ the quantitative analyses below). In contrast, in all higher eigenvalues, many level crossings appear, indicated by nonanalyticities (cusps) of any single level. The same findings hold also for quench protocol \circled{2} [panel (f)] with the difference that $\xi_0$ is of the same scale as $\xi_{r\ge1}$. Only at very early times after the quench, at $tJ \approx 0.9$, the entanglement spectrum becomes gapless, corresponding to a singularity in $g_r/g_1$ [cf.\ panel (g)]. 

The time dependence of the gap ratios $g_r/g_1$ is shown in panels (c) and (g) for $r = 2,\ldots,5$. We want to contrast their behavior to CFT expectations in case of quenches to the critical point (cf.\ App.\,\ref{app:critical}). In the latter case, the ratios assume the constant values $g_r/g_1 = \Delta_r/\Delta_1$, where $\Delta_r$ are the conformal dimensions of primary fields and their descendants in the boundary CFT \cite{Cardy:2016fqc}. In the ferromagnetic meson quench in panel (c), all shown values are instead oscillating at later times around values smaller than the lowest integer CFT value $g_2/g_1 = 3$ (indicated by the grey dashed line). In particular, $g_2/g_1$ (green curve) exhibits multiple nonanalytic cusps, when the gap between $\xi_1$ and $\xi_2$ closes and the ratio hence assumes the value one as the lower bound. The same features exist also under quench \circled{2} [panel (g)]. Here, the oscillations display a larger amplitude and assume higher values (shown on a logarithmic scale), while a single singularity appears immediately after the quench.\,\footnote{We refer to App.\,\ref{app:ratios_comparison} for a comparison of the ratio $g_2/g_1$ in all quench examples.} Since also higher order ratios exhibit cusps, when the gap between other eigenvalues closes, these gap ratios do not contain information on meson masses.

The behavior of the first four return rate functions $r_i$ is visible in panels (d) and (h). For quench protocol \circled{1}, $r_1$ is on a much smaller scale than all higher order ones. It exhibits regular oscillations, which carry the meson content of the post-quench Hamiltonian (cf.\ the discussions below). On the other side, all $r_i$ in quench type \circled{2} exhibit multiple level crossings. Since the first cusp in $r_1$ appears before the first minimum, we can identify them as \textit{regular} ones according to the nomenclature in \cite{PhysRevB.96.134427}.\,\footnote{While we do not observe any nonanalyticities in quench type (1), the existence of \textit{anomalous} cusps \cite{PhysRevB.96.134427} could be possible for other quench parameters within the ferromagnetic phase.} In contrast to the DQPT regime in the transverse Ising model (cf.\ Fig.~\ref{fig:spectra_transverse} in App.\,\ref{app:critical}), which is also characterized by regular cusps, they are, however, not equally spaced in time. Moreover, while regular cusp positions coincide in the previous case with times when the entanglement spectrum becomes gapless \cite{Torlai_2014}, this is not a necessary consequence in the mesonic regime under consideration, i.e.\ the modular Hamiltonian remains gapped at these points in time, apart from the single exception at early times.

\change{We use different methods in this letter to analyze the meson content of entanglement spectra quantitatively and draw reliable interpretations from them. Fig.\,\ref{fig:Prony_type12} shows the results of a Prony signal analysis, whose basic idea is to represent a function as a sum of complex exponentials with frequencies plotted in the complex plane (see App.\,\ref{app:Prony} for more details).} 
The first row displays the analysis of $S_1$ in comparison to $\xi_0$ in the second row. In quench type \circled{1} [panels (a,b)] within the ferromagnetic phase, both quantities allow the clear \change{and stable} detection of four meson states $M_i$, which are consistent with a semiclassical approximation \cite{Kormos2017} (shown as grey vertical lines). Additionally, meson mass differences $m_{ij}$ (shown as green vertical lines) and the continuum threshold at $2 M_1$ can be identified. When the initial state is in the paramagnetic phase, i.e.\ for type \circled{2} [panels (c,d)], remnants of the meson states are still visible, but less clear due to the large entanglement growth. In both quenches, one can observe that $\xi_0$ even allows for a clearer extraction of meson poles in the Prony analyses than $S_1$. These analyses show that the meson content of the post-quench Hamiltonian, giving rise to entanglement oscillations, is fully encoded in the dominant eigenvalue of the modular Hamiltonian.\,\footnote{All higher order eigenvalues and gaps exhibit nonanalytic cusps in their profile and hence do not allow to define oscillation frequencies, which could be quantitatively analyzed.} Interestingly, $r_1$ in quench type \circled{1} equally encodes the meson masses in the frequency pattern, but in contrast to $\xi_0$, neither mass differences, nor the continuum threshold are appearing.

\section{Quenches to the integrable E$_8$ QFT regime}

\begin{figure}[t]
    \centering
   \includegraphics[width=0.49\columnwidth]{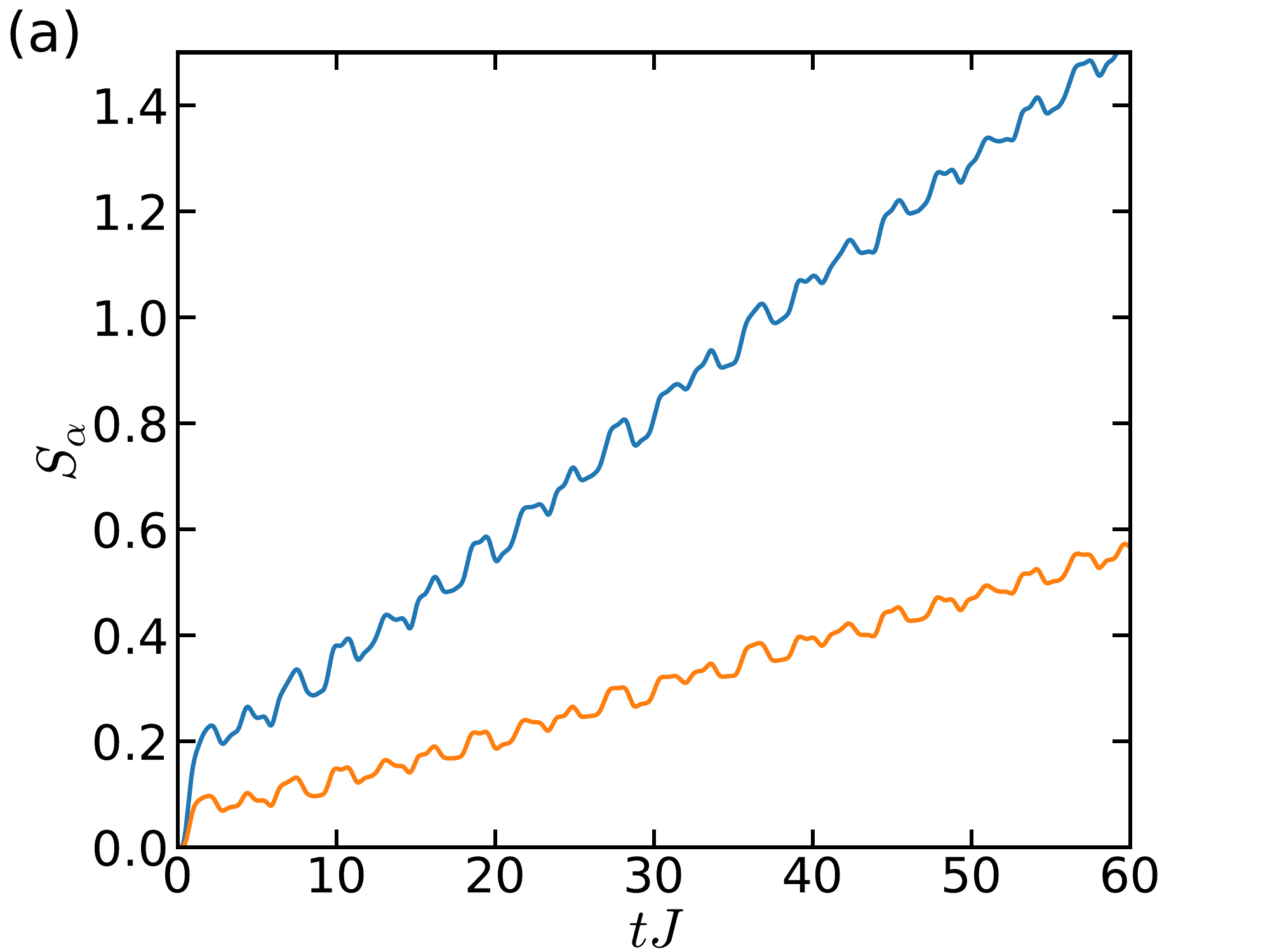} 
   \includegraphics[width=0.49\columnwidth]{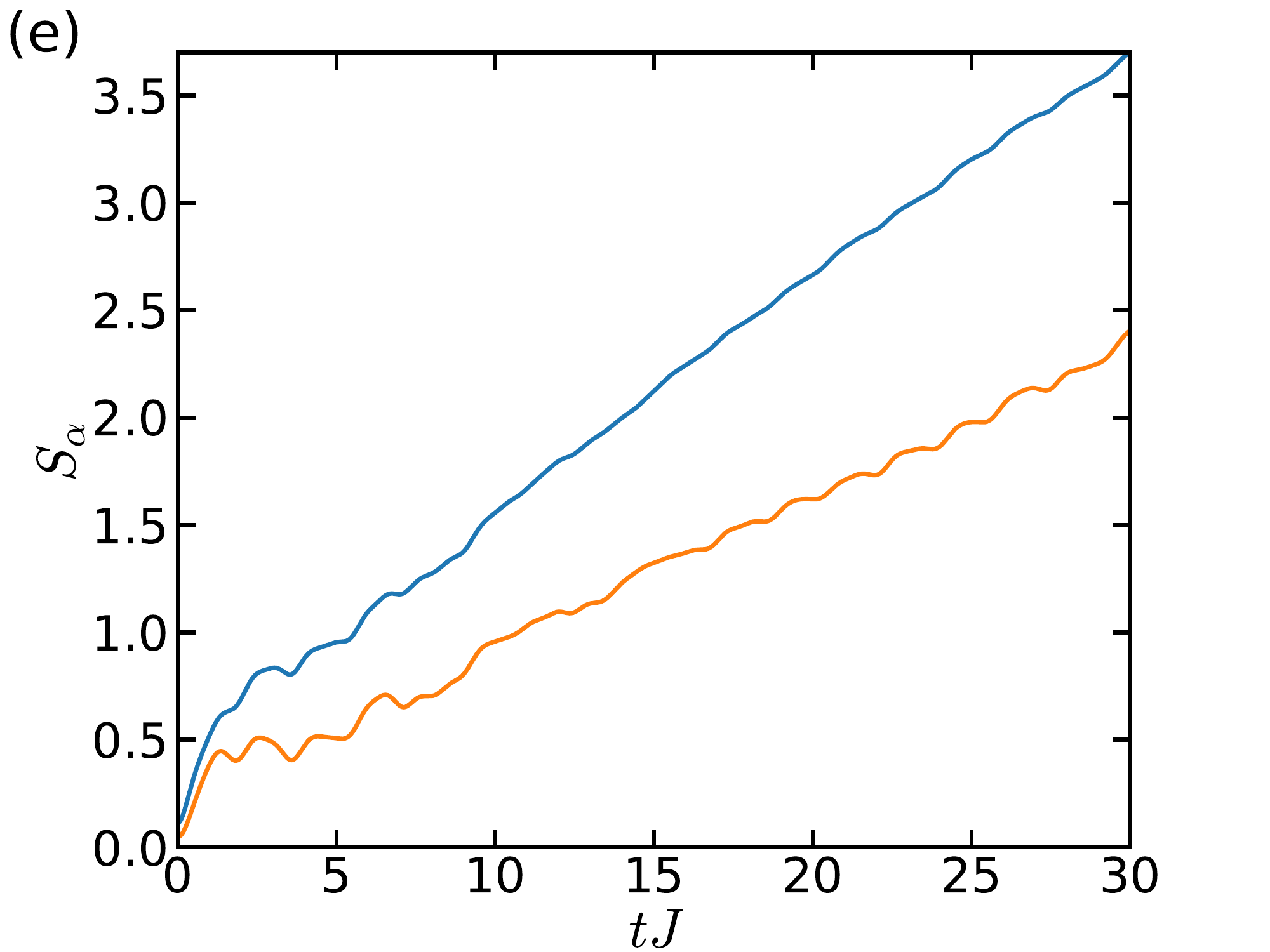} 
   \includegraphics[width=0.49\columnwidth]{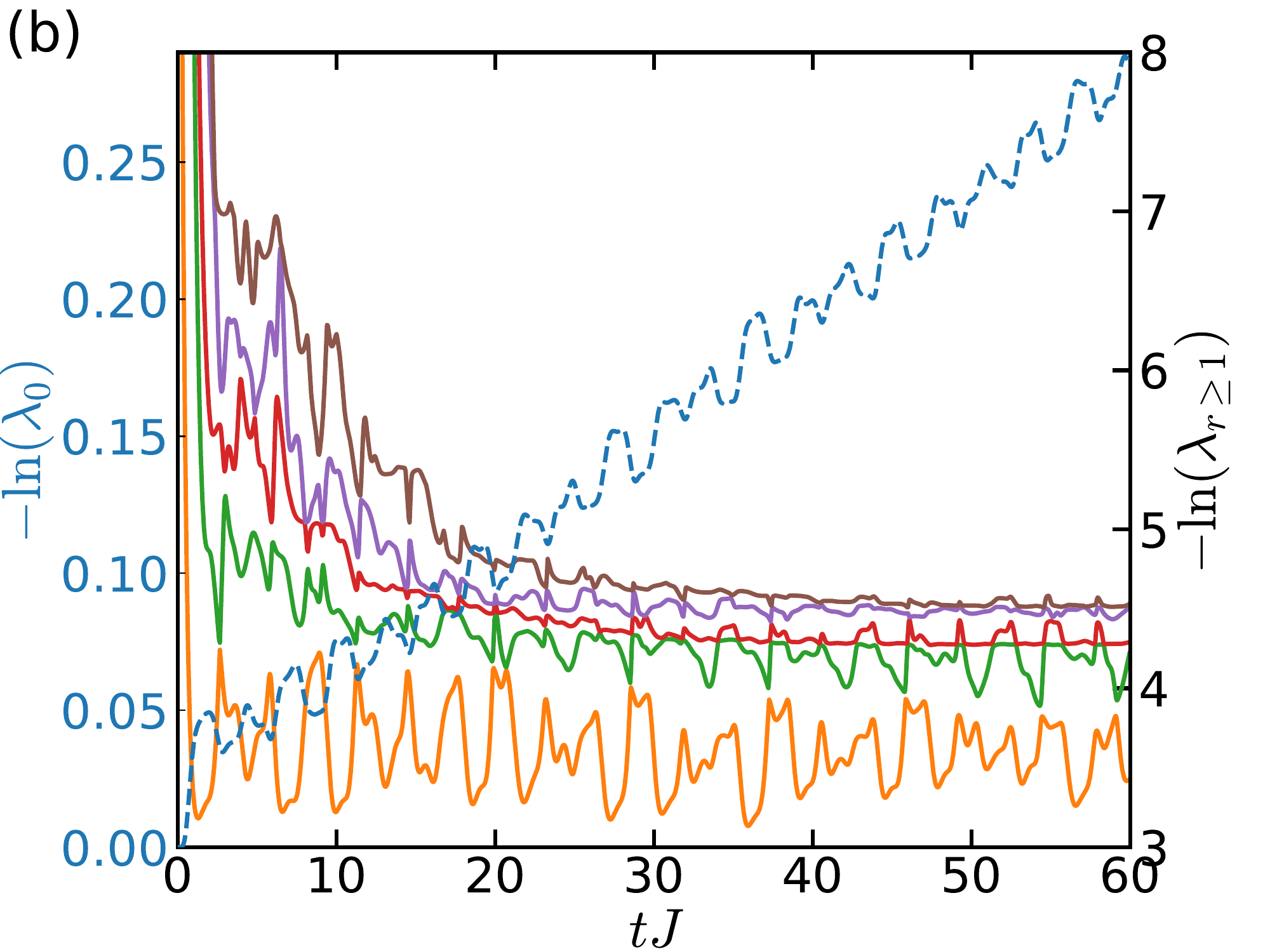} 
   \includegraphics[width=0.49\columnwidth]{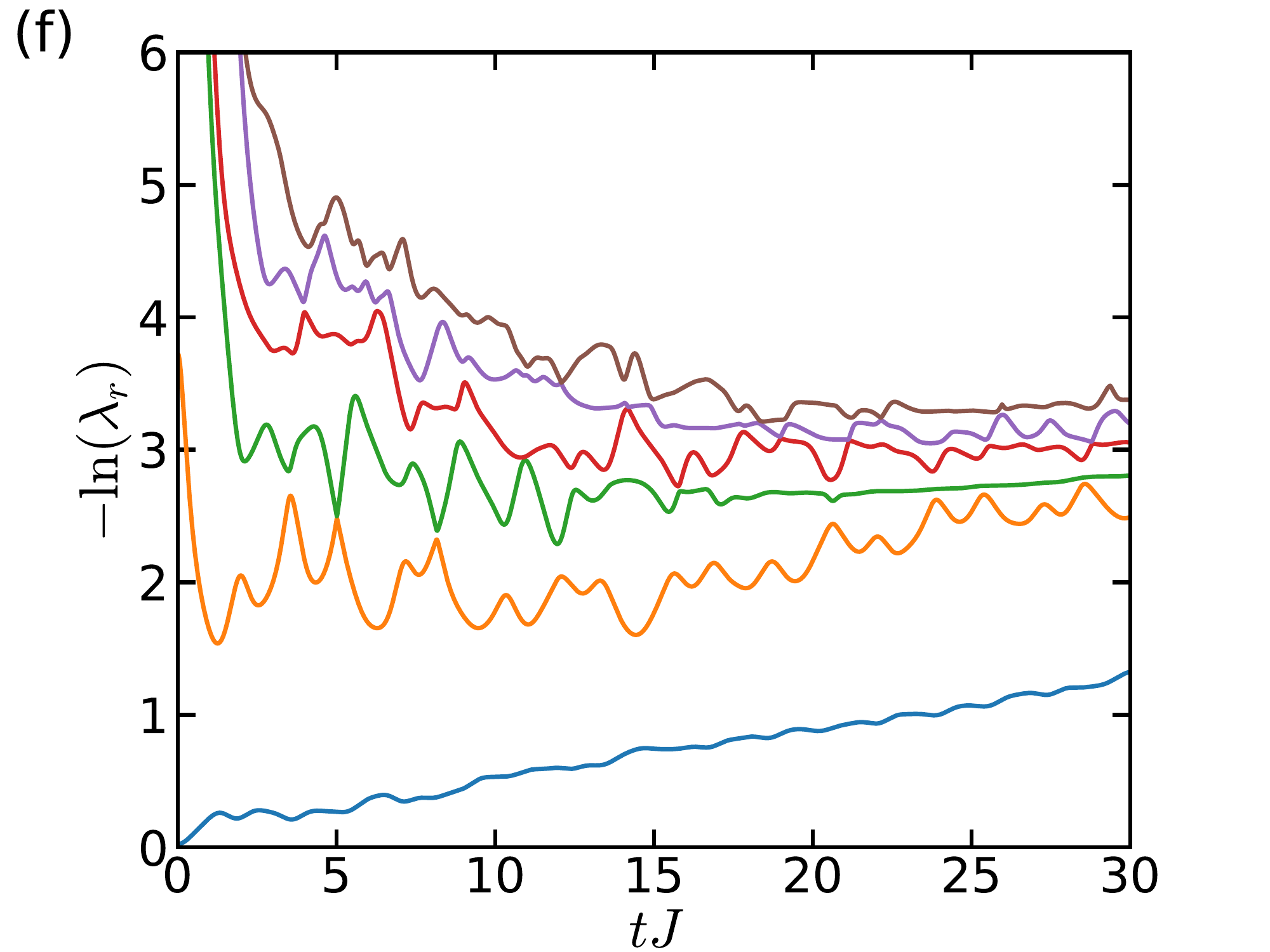} 
   \includegraphics[width=0.49\columnwidth]{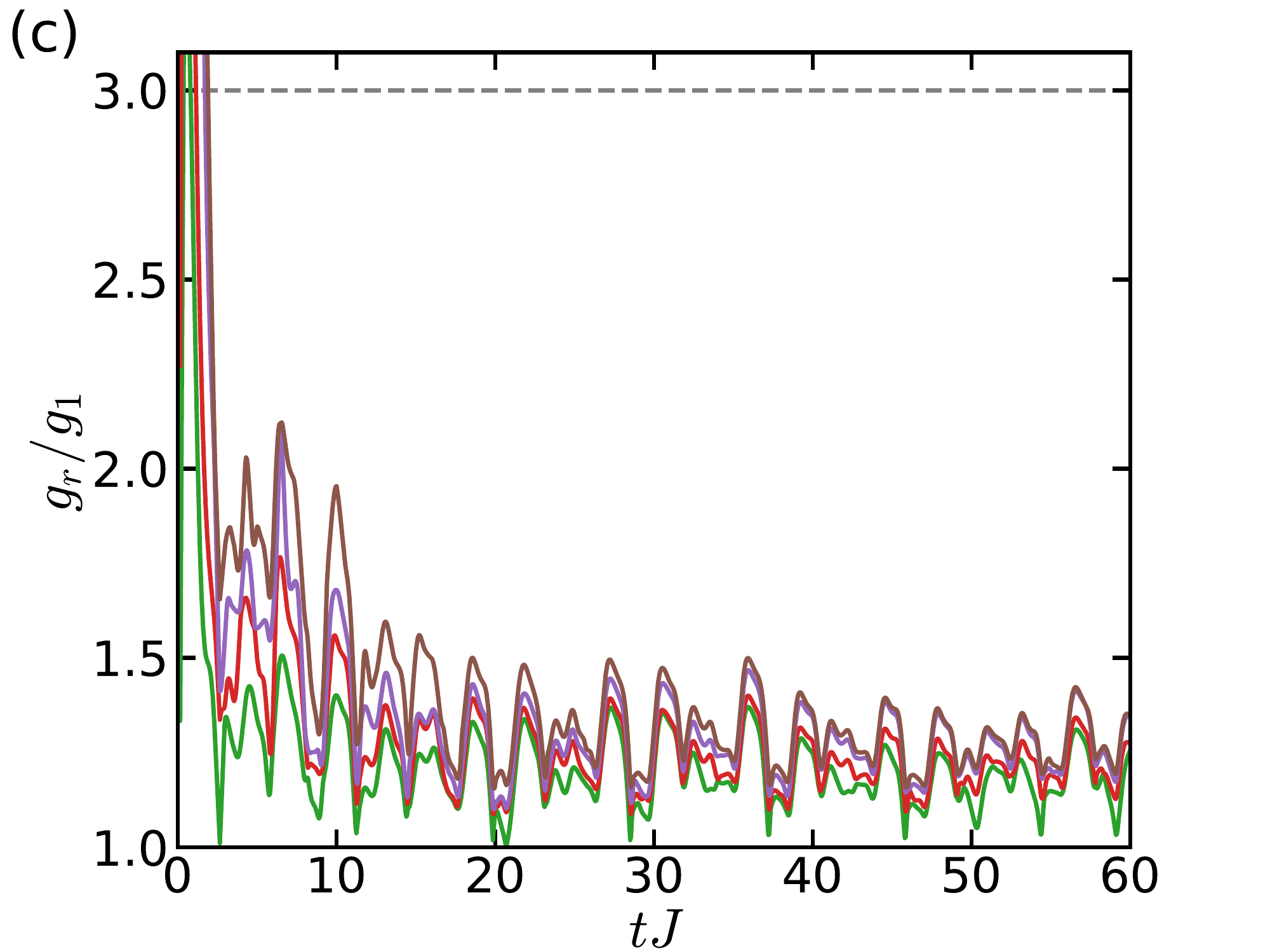} 
   \includegraphics[width=0.49\columnwidth]{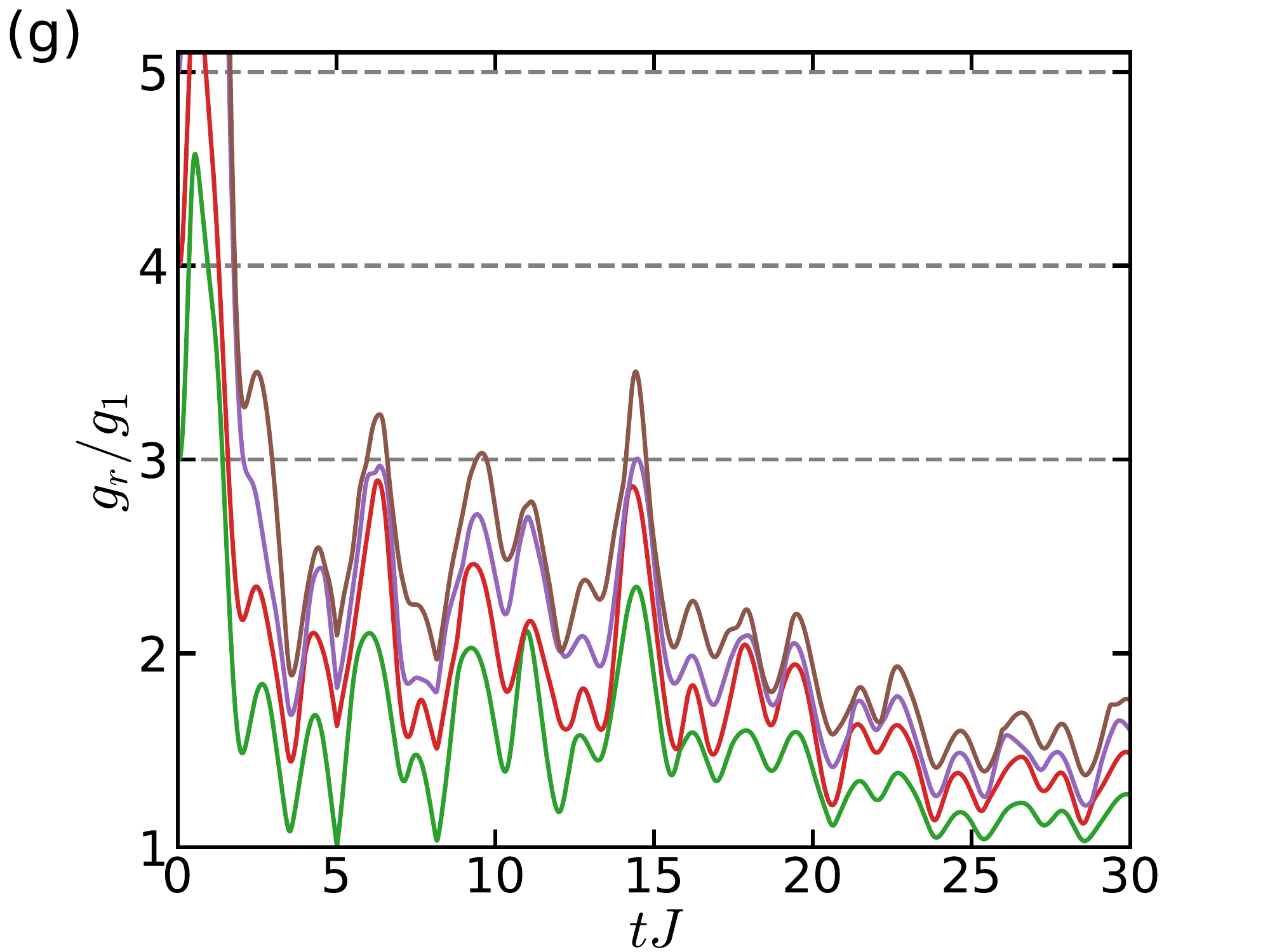} 
   \begin{minipage}{0.49\columnwidth}
   \includegraphics[width=\textwidth]{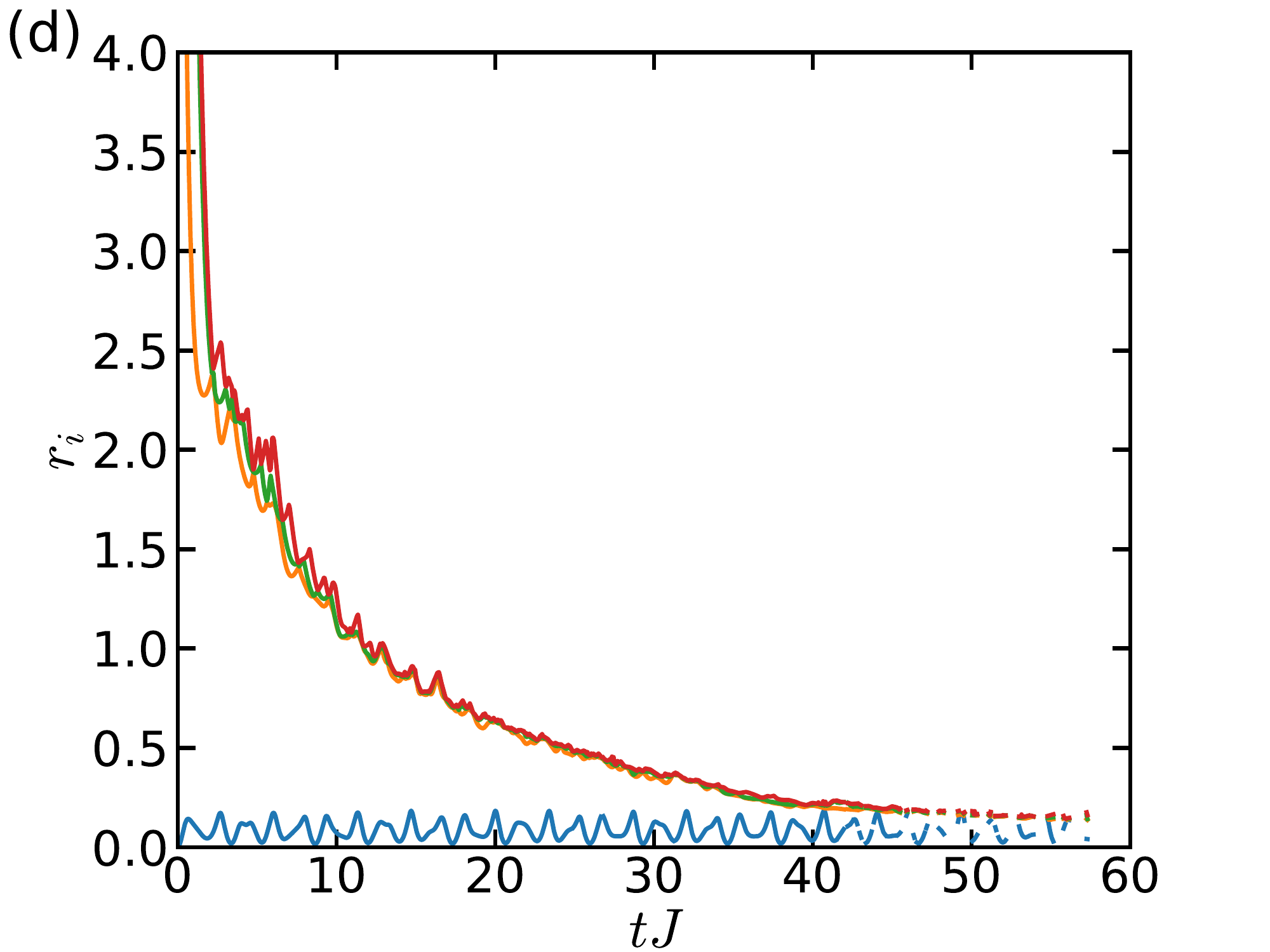} \\ 
   \circled{3}
   \end{minipage}
   \begin{minipage}{0.49\columnwidth}
   \includegraphics[width=\textwidth]{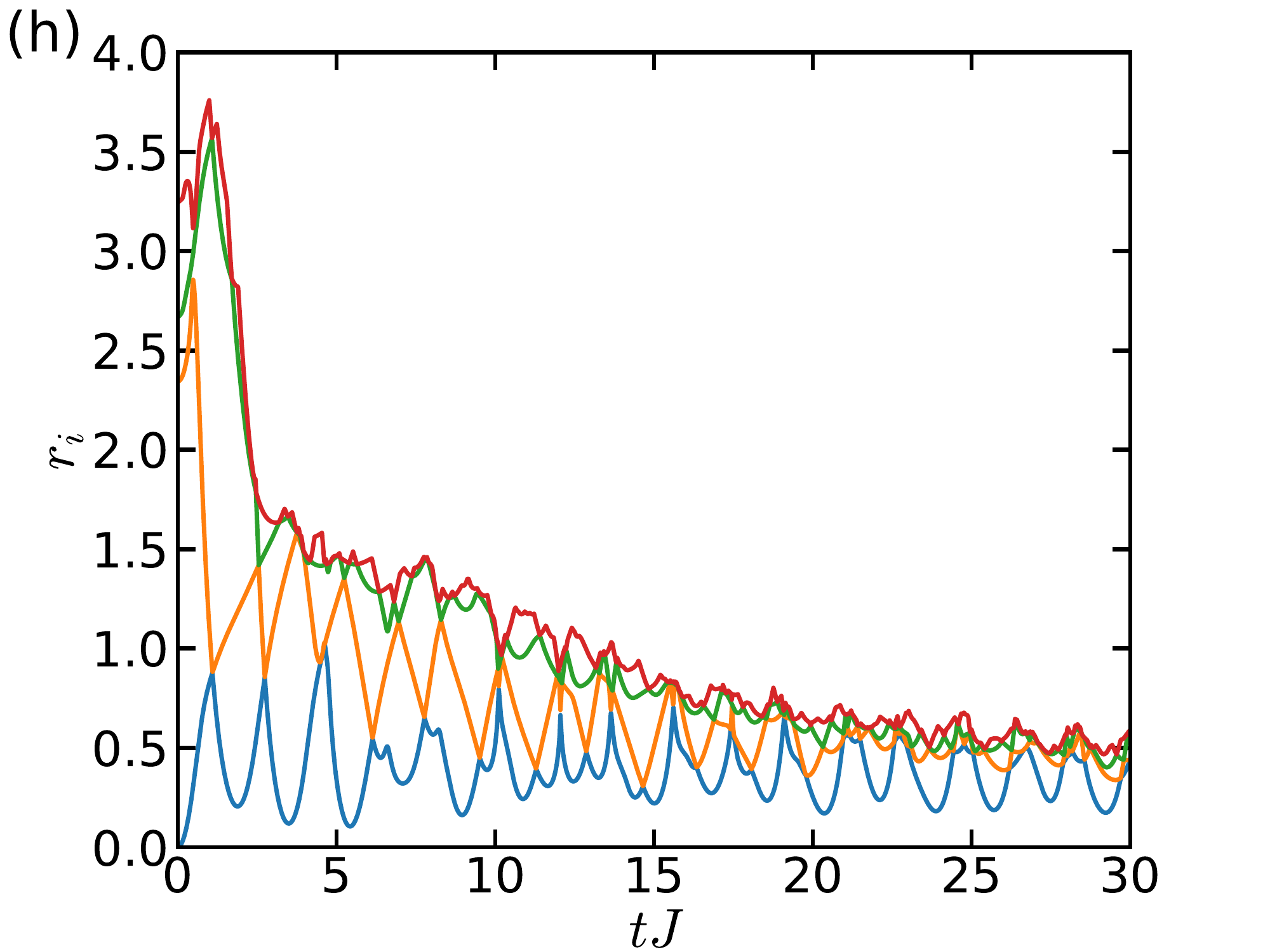} \\ 
   \circled{4}
   \end{minipage}
    \caption{Time dependence of physical quantities in quench protocols (3) (a-d) and (4) (e-h) to the integrable E$_8$ QFT regime. Legends and quantities are the same as in Fig.\,\ref{fig:spectra_meson}. See text for detailed discussions.}
    \label{fig:spectra_E8}
\end{figure}
\begin{figure}[t]
    \centering
   \includegraphics[width=0.49\columnwidth]{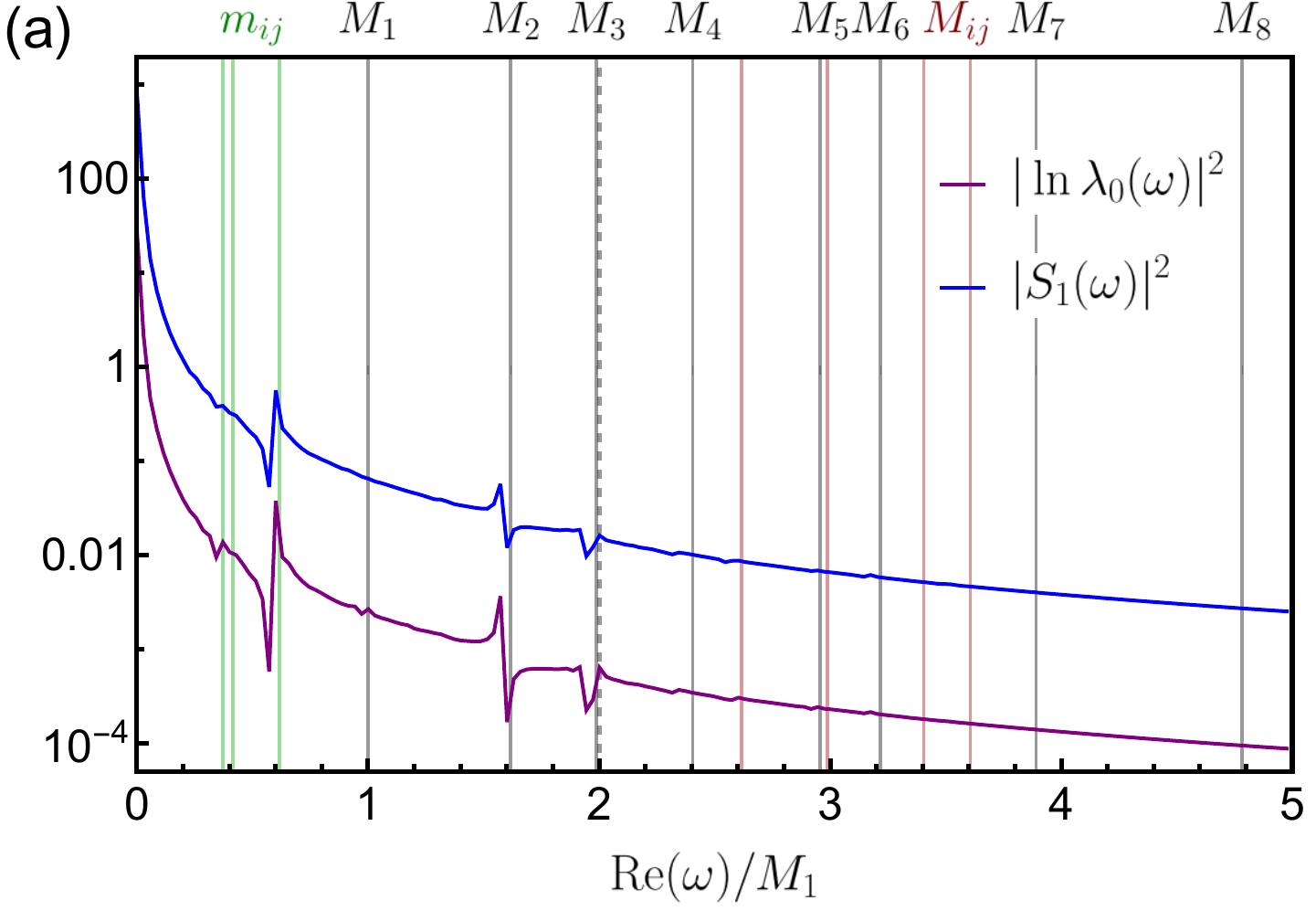} 
   \includegraphics[width=0.49\columnwidth]{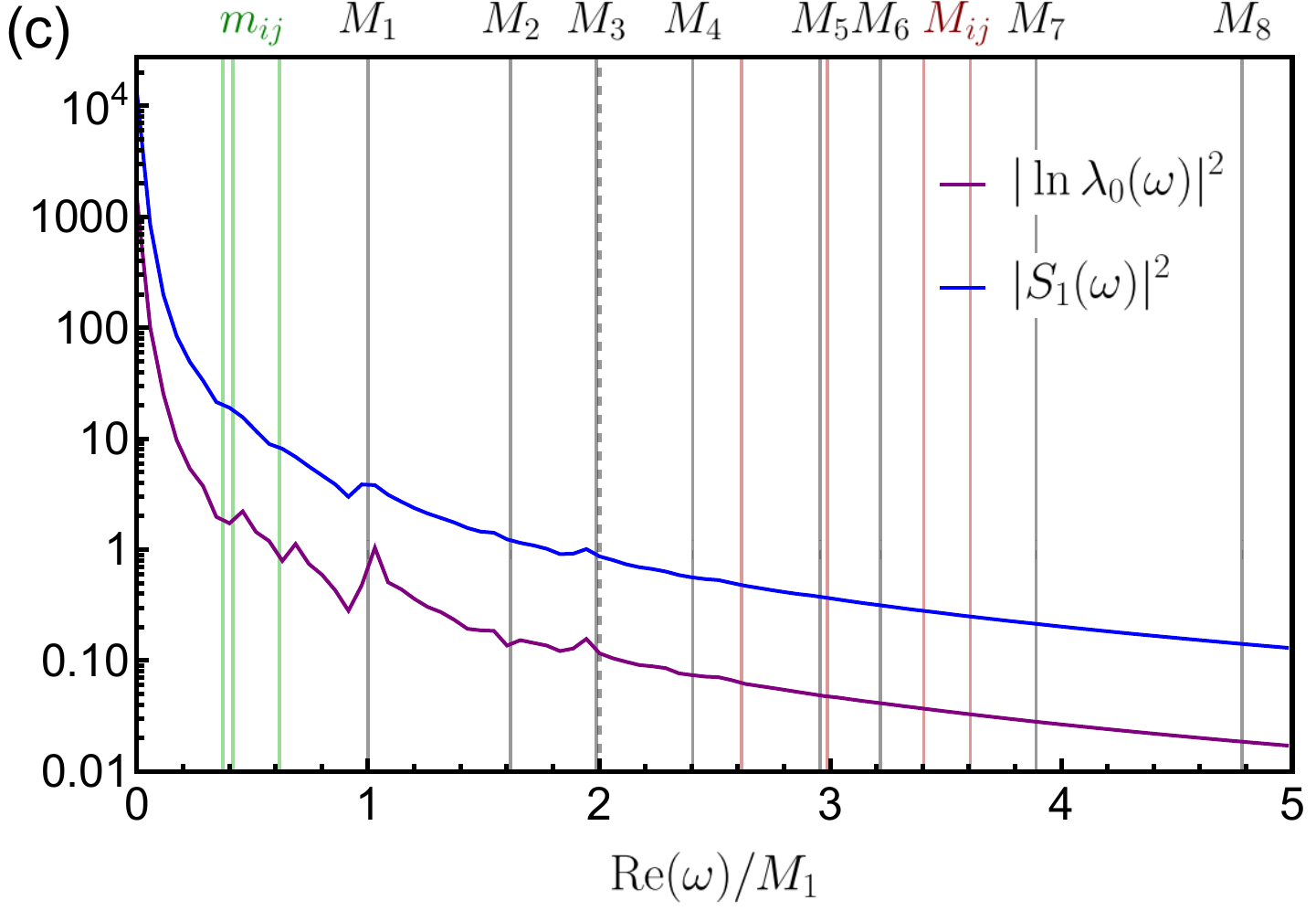} 
   \begin{minipage}{0.49\columnwidth}
   \includegraphics[width=\textwidth]{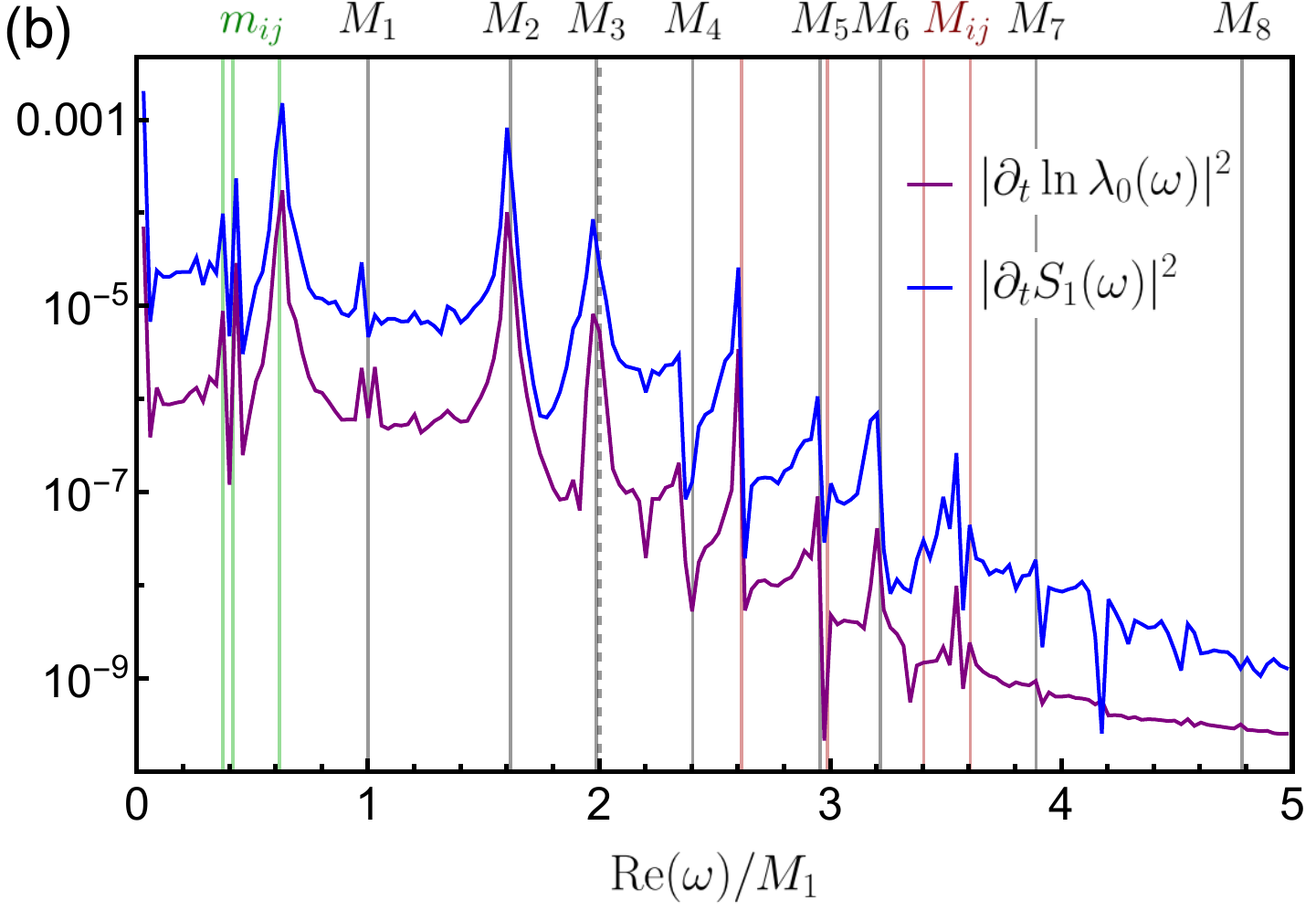}\\ 
   \circled{3}
   \end{minipage}
   \begin{minipage}{0.49\columnwidth}
   \includegraphics[width=\textwidth]{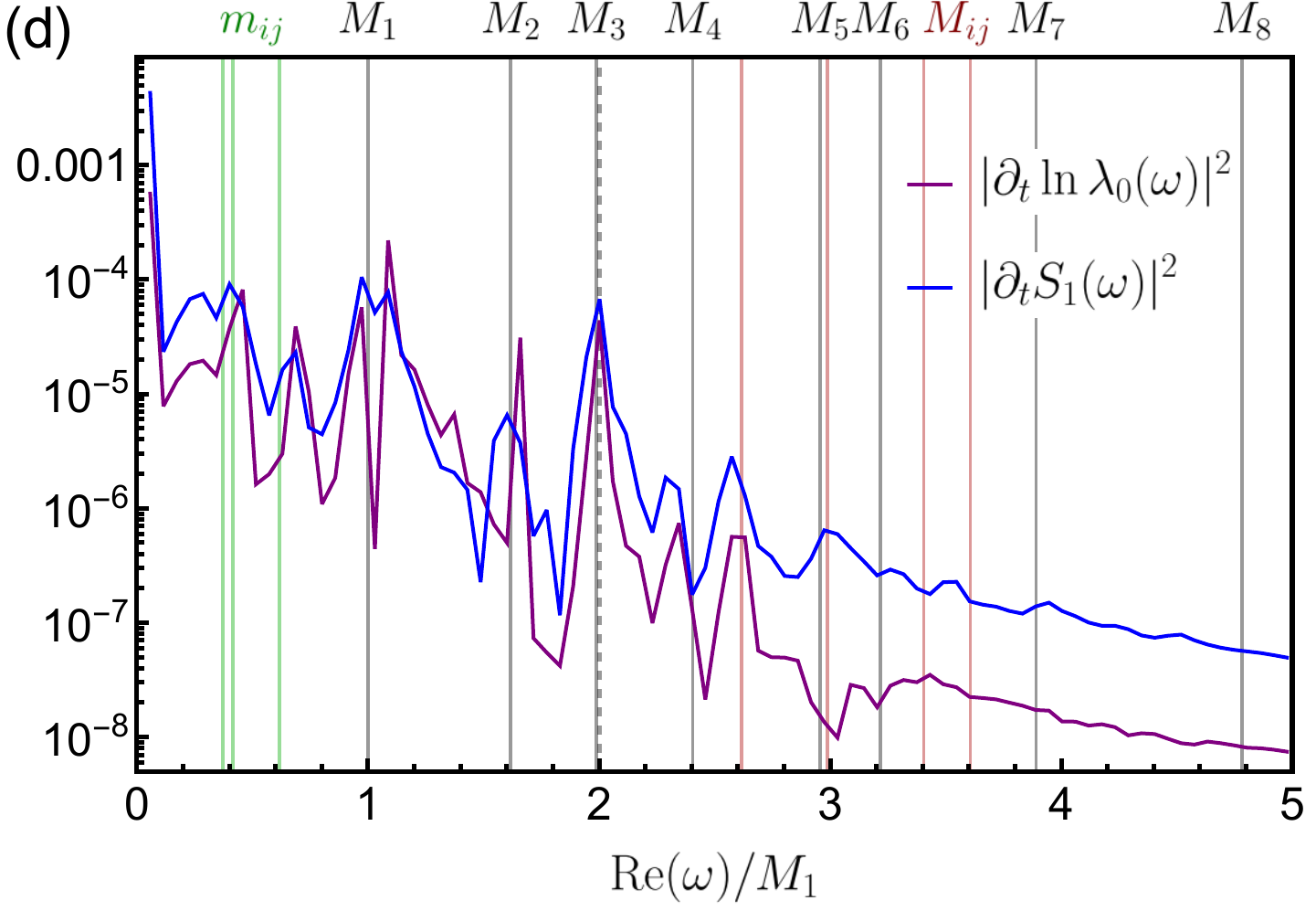}\\ 
   \circled{4}
   \end{minipage}
    \caption{Fourier spectra of $\xi_0$ and $S_1$ (a,c), and their time derivatives (b,d) under quench type (3) (left column) and (4) (right column). Green background lines mark the following mass differences: $m_{23}$, $m_{34}$, $m_{12}$ (ascending). Red vertical lines indicate the following mass sums $M_{ij}\equiv M_i+M_j$: $M_{12}$, $M_{13}$, $M_{14}$, $M_{23}$ (ascending). 
    The results allow to identify several meson states equally accurate from entanglement oscillations in $S_1$ and $\xi_0$.}
    \label{fig:FT_type34}
\end{figure}

We now consider protocols \circled{3} and \circled{4}, which quench towards the integrable E$_8$ QFT regime with 8 stable meson states. Fig.\,\ref{fig:spectra_E8} shows the simulation results. In type \circled{3} [panel (a)], $S_1$ and $S_2$ show entanglement oscillations, which, in contrast to \circled{1}, are not bounded but superimposed with a linear growth. As discussed e.g.\ in \cite{Birnkammer:2022giy,Scopa:2021gcx}, such a behavior can be explained in a quasiparticle picture by mesons produced at finite velocity (due to a large quench magnitude), which are able to spread entanglement and quantum correlations faster. As in the previous section, the entanglement oscillations in the same quantities are much less pronounced under quench type \circled{4} [panel (e)], when the initial state is in the paramagnetic phase.

The corresponding entanglement spectra [panels (b,f)] are gapped. As in the semiclassical regime, $\xi_0$ (blue curves) shares the qualitative behavior of $S_1$ in both quenches. Similarly, multiple level crossing appear in all higher order eigenvalues. The gap ratios $g_r/g_1$ [panels (c,g)] are oscillating around lower values than the constant CFT value. For $g_2/g_1$, the oscillations are broken by several cusps at the minimal lower value. While the return rate density $r_1$ in type \circled{3} shows oscillations, which are given by the E$_8$ meson masses, only higher order rate functions $r_{i>1}$ exhibit level crossings [panel (d)].\,\footnote{In the calculation of $r_i$, some numerical exception errors occur at late times. The corresponding data points are left blank in panel (d).} The same quantity in quench type \circled{4} [panel (h)] instead has numerous regular cusps at unequally spaced positions, indicating the appearance of DQPTs.

Fig.\,\ref{fig:FT_type34} shows a Fourier analysis of $S_1$ (blue curves) and $\xi_0$ (purple curves) for both quenches [panels (a,c)]. Due to the dominating linear entanglement growth, the Fourier spectra are decreasing towards larger frequencies and overall relatively flat with only small peak structures. For that reason we evaluate in panels (b,d) also their time derivatives, which allow to identify the oscillating contributions more clearly. Several peaks become discernible that match the analytical E$_8$ meson mass ratios as well as some mass differences and sums. For quench \circled{4} from the paramagnetic phase, these features are much less pronounced. There are only mild differences between the behavior of $S_1$ and $\xi_0$.

The discussions of this section exemplify that the previously found conclusions in the semiclassical meson regime hold equally also in the relativistic E$_8$ QFT. That is, the dominant eigenvalue of the entanglement spectrum fully encodes the meson content of the QMB system or QFT. The appearance of regulars cusps at irregular positions, indicating the appearance of DQPTs, does not imply that the entanglement spectrum becomes gapless at these points in time.

\section{Discussion and outlook}

In this letter we have studied the impact of meson confinement on the time evolution of the lowest eigenvalues in the entanglement spectrum and return rate functions after global quantum quenches in the Ising model. Our analyses contribute to a deeper understanding of entanglement properties of emergent phenomena in QMB systems and QFTs. The study of meson confinement and DQPT properties in the (1+1)-dimensional Ising QFT is a first step to more complex systems akin to QCD in particle physics. This necessarily involves the consideration of gauge theories, where the existence of DQPTs has been predicted in \cite{Huang:2018vss,Zache:2018cqq} and further investigated in \cite{VanDamme:2022loc,VanDamme:2022rnb}. Very recently, its first experimental observation was realized on a quantum computer and simultaneously discussed with entanglement tomography \cite{Mueller:2022xbg}. As a key implication of our study we see the potential use of such tomographic experiments to access the meson content of entanglement oscillations from the lowest part of the entanglement spectrum, instead of the experimentally inaccessible entanglement entropy itself.

\begin{acknowledgments}
We would like to thank Mari Carmen Ba\~nuls, Jens Eisert, Jad C.\ Halimeh, Michal P.\ Heller, Jacopo Surace and Luca Tagliacozzo for useful discussions and correspondence. JK is supported by the Israel Academy of Sciences and Humanities \& Council for Higher Education Excellence Fellowship Program for International Postdoctoral Researchers.
\end{acknowledgments}

\bibliographystyle{jk_ref_layout_wTitle} 
\bibliography{literature}

\clearpage
\appendix
\pagenumbering{gobble}
\setcounter{equation}{0}


\onecolumngrid
\begin{center}
\textbf{Supplemental Material to\\
``Meson content of entanglement spectra after integrable and nonintegrable quantum quenches''}
~\\~\\
Johannes Knaute
\end{center}
\vspace{1cm}
\twocolumngrid

This supplemental material contains appendices, in which we discuss some further background material of our analyses.

\section{Quenches at criticality and in the massive free fermion regime}
\label{app:critical}

\begin{figure}[b]
    \centering
   \includegraphics[width=0.49\columnwidth]{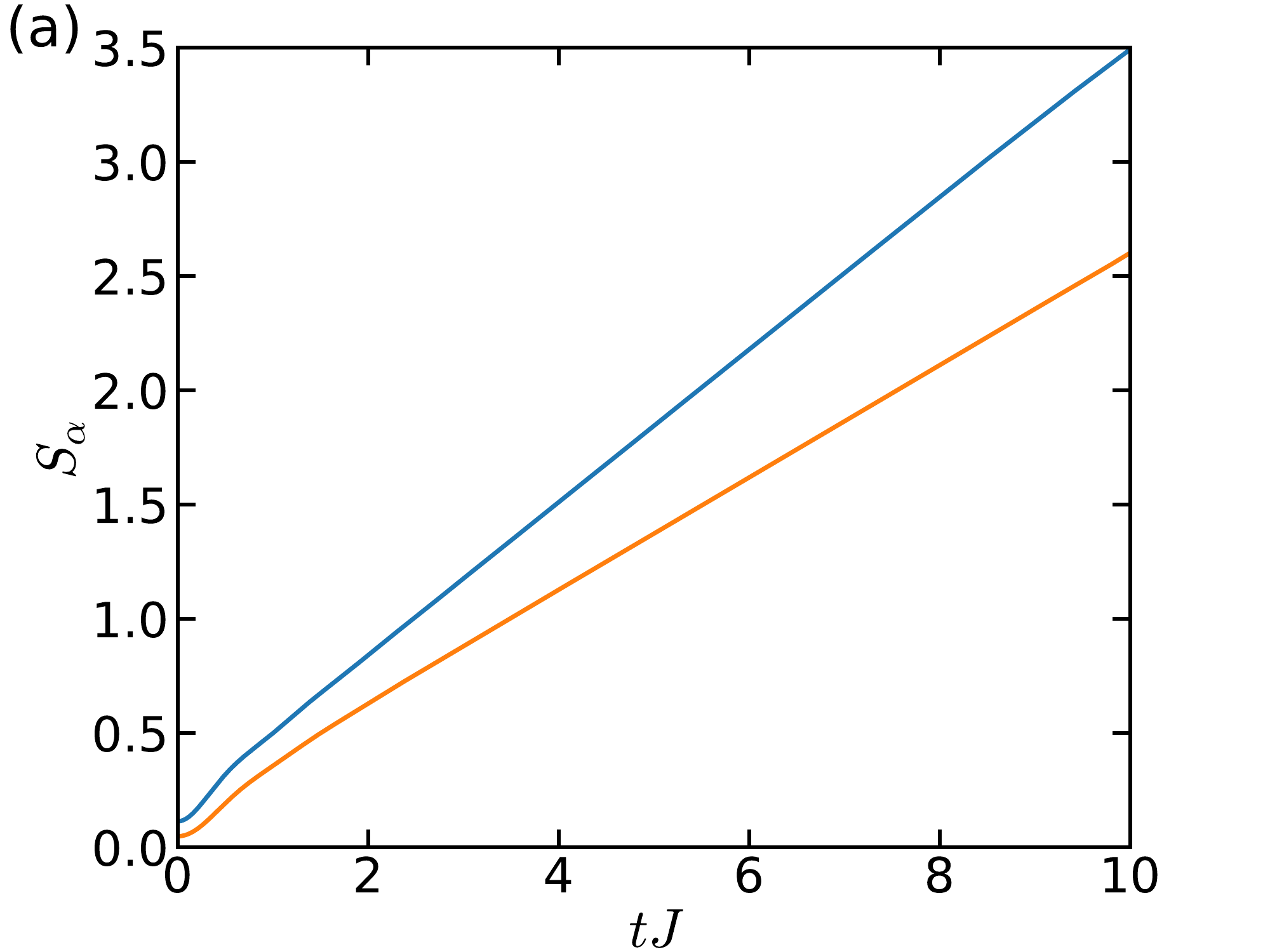}
   \includegraphics[width=0.49\columnwidth]{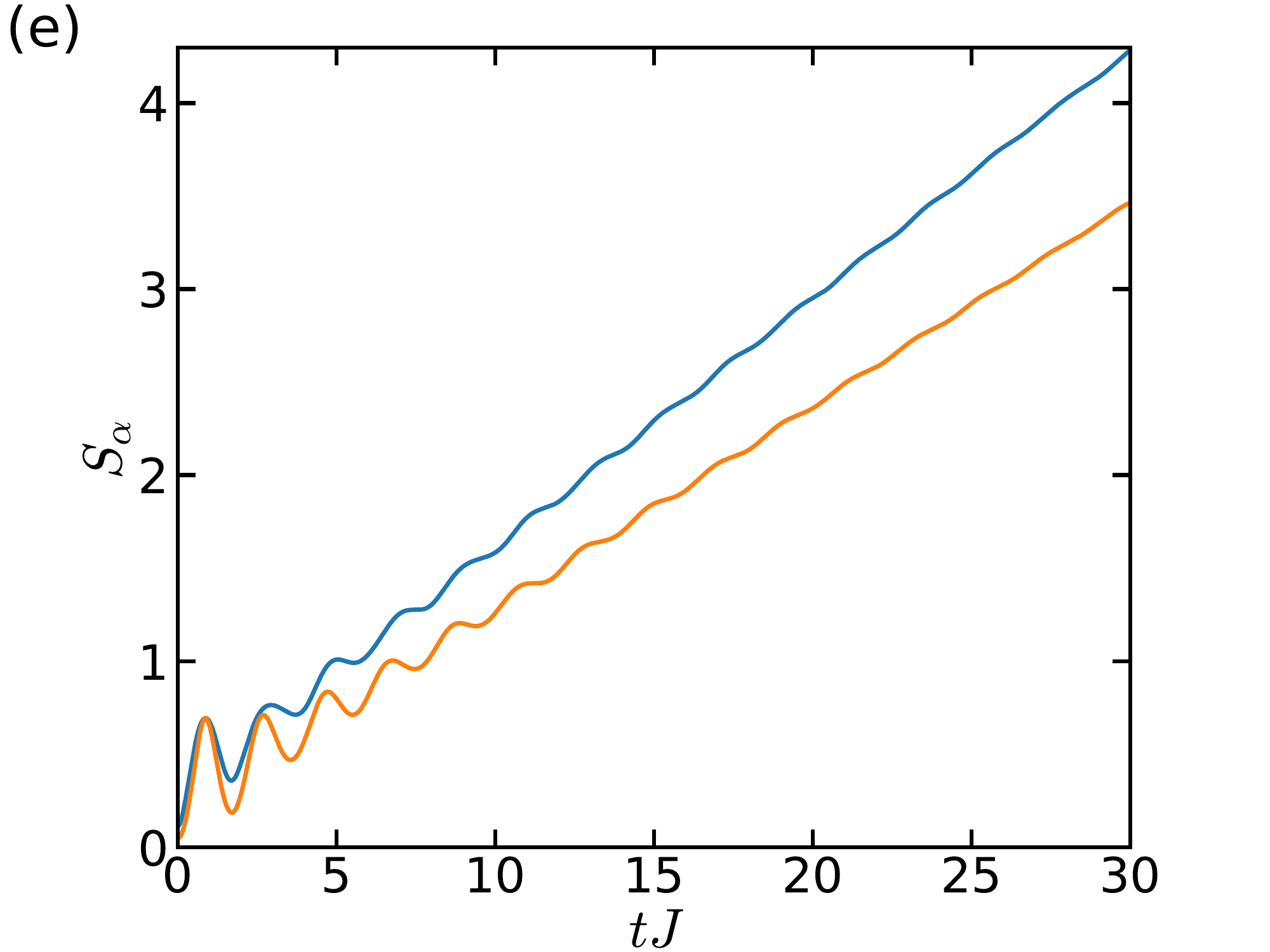}
   \includegraphics[width=0.49\columnwidth]{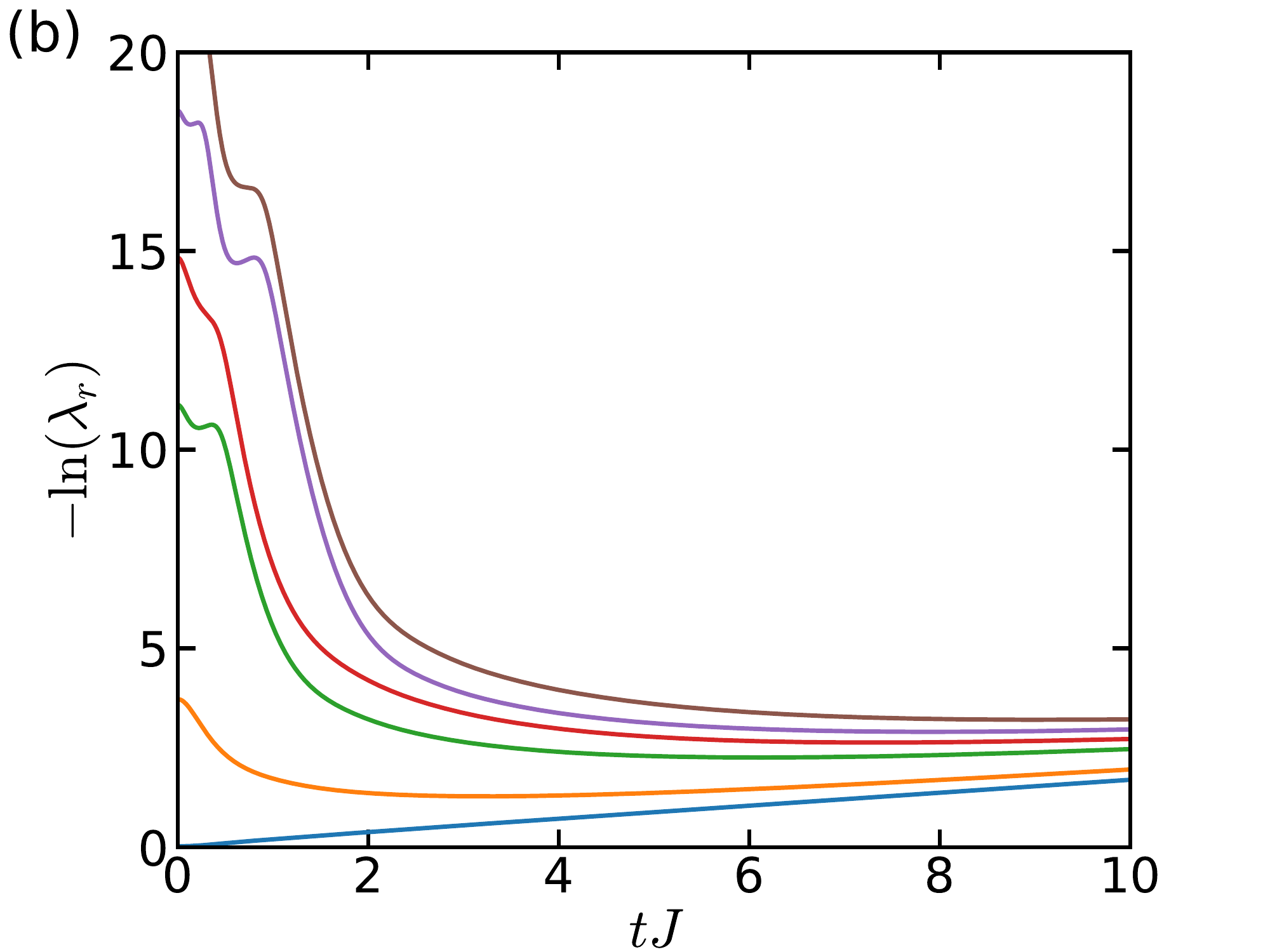}
   \includegraphics[width=0.49\columnwidth]{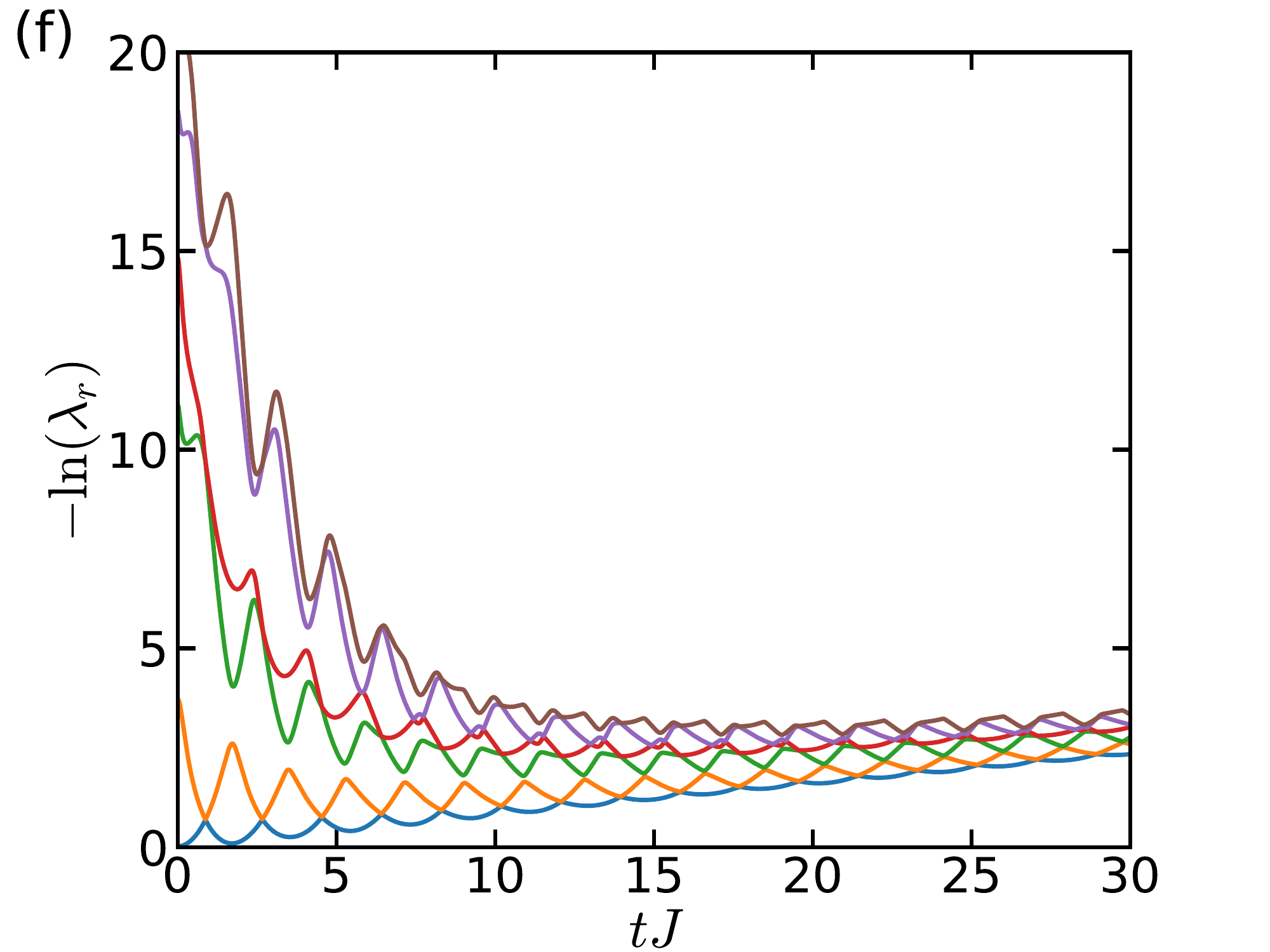}
   \includegraphics[width=0.49\columnwidth]{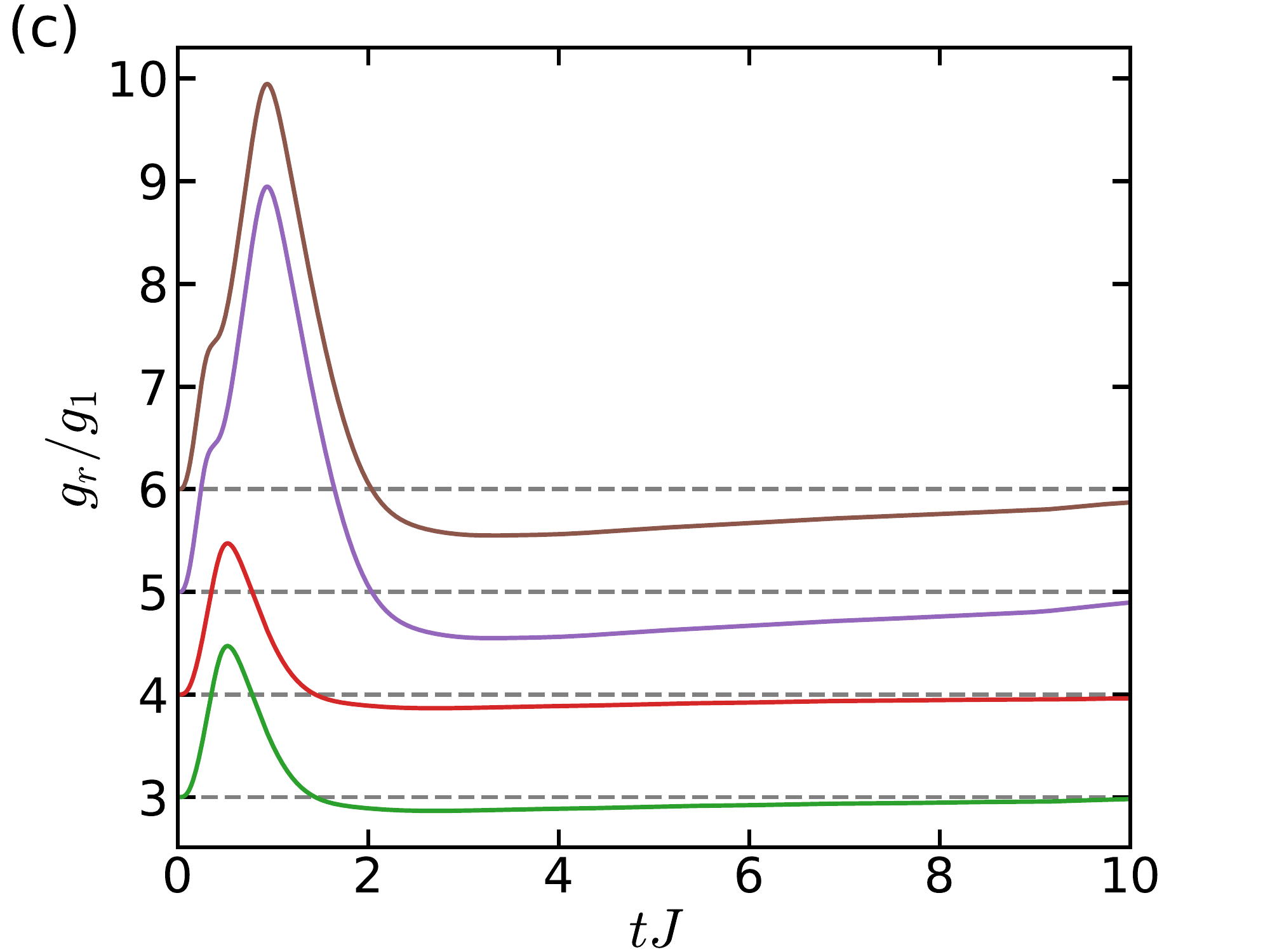}
   \includegraphics[width=0.49\columnwidth]{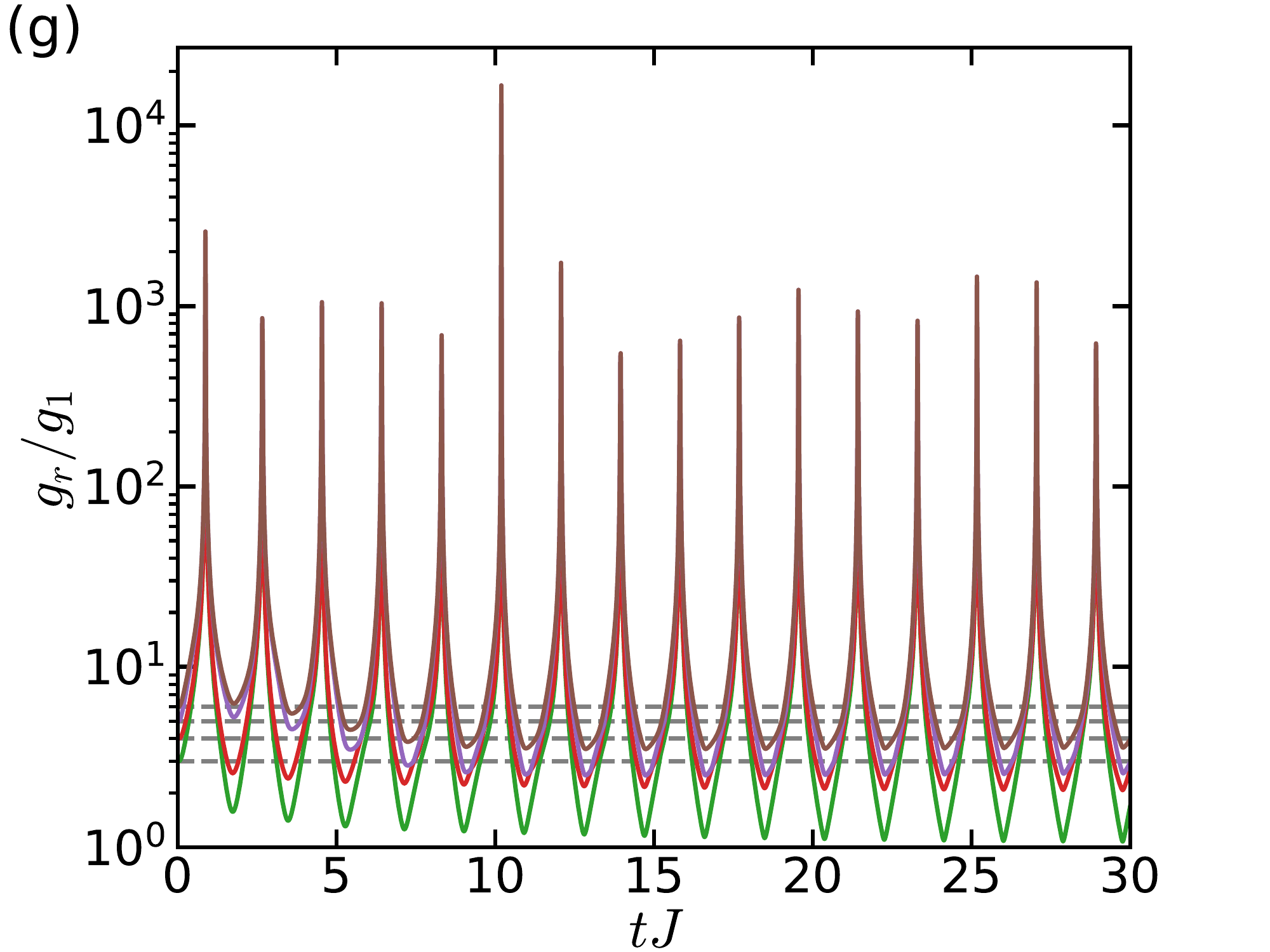}
   \begin{minipage}{0.49\columnwidth}
   \includegraphics[width=\textwidth]{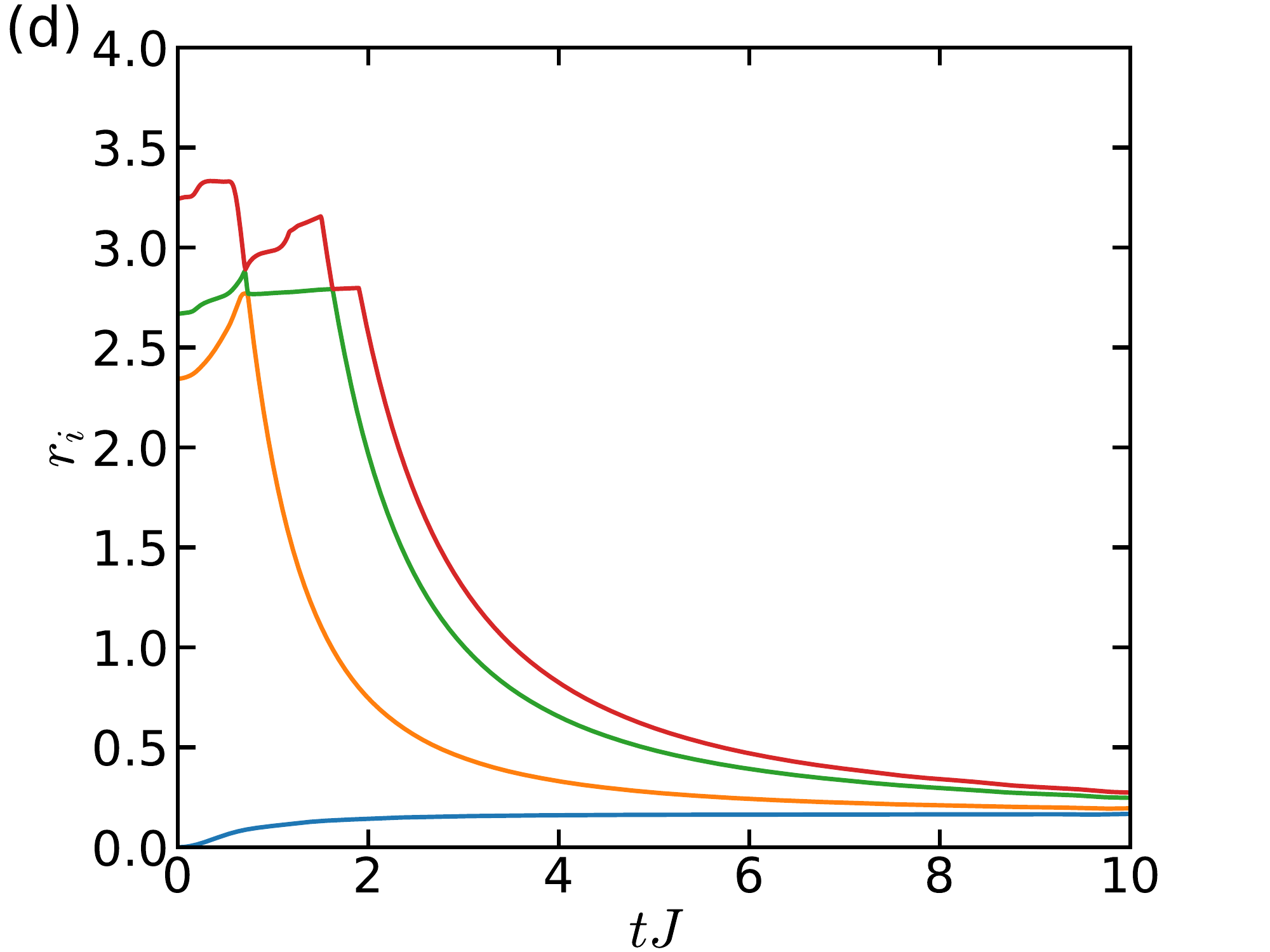} \\
   \circled{5}
   \end{minipage}
   \begin{minipage}{0.49\columnwidth}
   \includegraphics[width=\textwidth]{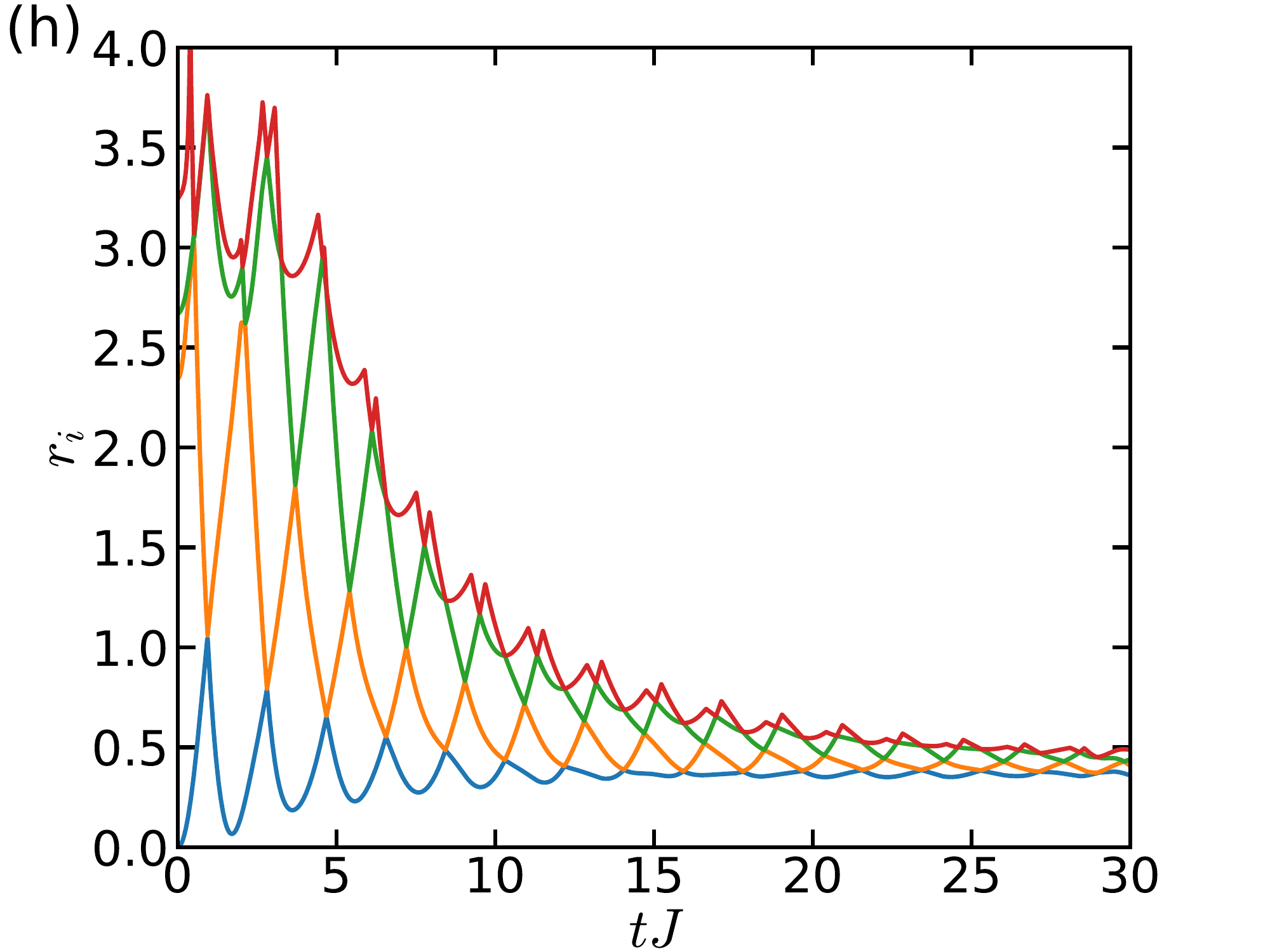} \\
   \circled{6}
   \end{minipage}
    \caption{Time dependence of physical quantities in quench protocol (5) (a-d) to the critical point and (6) (e-h) to the massive free fermion regime regime. Legends and quantities are the same as in Fig.\,2 of the main text.}
    \label{fig:spectra_transverse}
\end{figure}

In this appendix we revisit quenches in the transverse Ising model as a comparison to the mesonic cases studied in the main text. Fig.\,\ref{fig:spectra_transverse} shows the simulation results of quench protocol \circled{5} from the paramagnetic phase to the critical point, and protocol \circled{6} to the ferromagnetic phase (i.e.\ in the massive free fermion regime; cf.\ Fig.\,1). In type \circled{5}, $S_1$ and $S_2$ [panel (a)] exhibit an linear growth (after a short initial quench phase), which is consistent with expectations of quasiparticle model interpretations \cite{Calabrese:2005in}. As alluded in the main text, the entanglement gap ratios $g_r/g_1$ in panel (c), following from the entanglement spectrum in panel (b), are expected to carry universal information by assuming integer values related to the conformal dimensions in the boundary CFT. This observation was made originally in \cite{Cardy:2016fqc} for a semi-infinite bipartition, as we consider it here, but holds also for finite size blocks \cite{Surace:2019mft}. One can observe that the lowest two ratios assume their corresponding values (shown as gray dashed lines) very accurately already at early times. Higher order eigenvalues deviate more and tend to converge to the analytical ratios towards later times. Note that these curves are valid for the particular initial state constructed in the main text (with a transverse field value $h=1.75$). In general, one can observe that the analytical ratios are approached more accurately, the more the initial state is in the paramagnetic phase, i.e.\ for $h \gg 1$. There is a gap in panel (d) between the monotonously increasing return rate density $r_1$ and higher order rate functions $r_{i>1}$.

In contrast, under quench type \circled{6}, many level crossing appear in the rate functions $r_i$, indicated by equally spaced regular cusps appearing in the time evolution of $r_1$; cf.\ panel (h). At these points in time, the entanglement spectrum becomes gapless [cf.\ panel (f)], such that the gap ratios $g_r/g_1$ diverge (cf.\ panel (g) on a logarithmic scale).

\section{Details on the iTEBD simulations}
\label{app:iTEBD}

By monitoring the truncation error in the iTEBD simulations for sufficiently high bond dimensions $D$, we estimate the reachable time scales in the different quench scenarios. More concretely, we use a dynamical truncation, where singular values below machine precision are discarded as long as $D$ is below its maximal value. For quench type (1) with bounded entanglement growth, $D=16$ then suffices. In all other protocols, we use the maximal value $D=500$, while the truncation error never exceeds $10^{-7}$ at late times.

\section{Entanglement gap ratios in mesonic quenches}
\label{app:ratios_comparison}

\begin{figure}[h]
    \centering
   \includegraphics[width=\columnwidth]{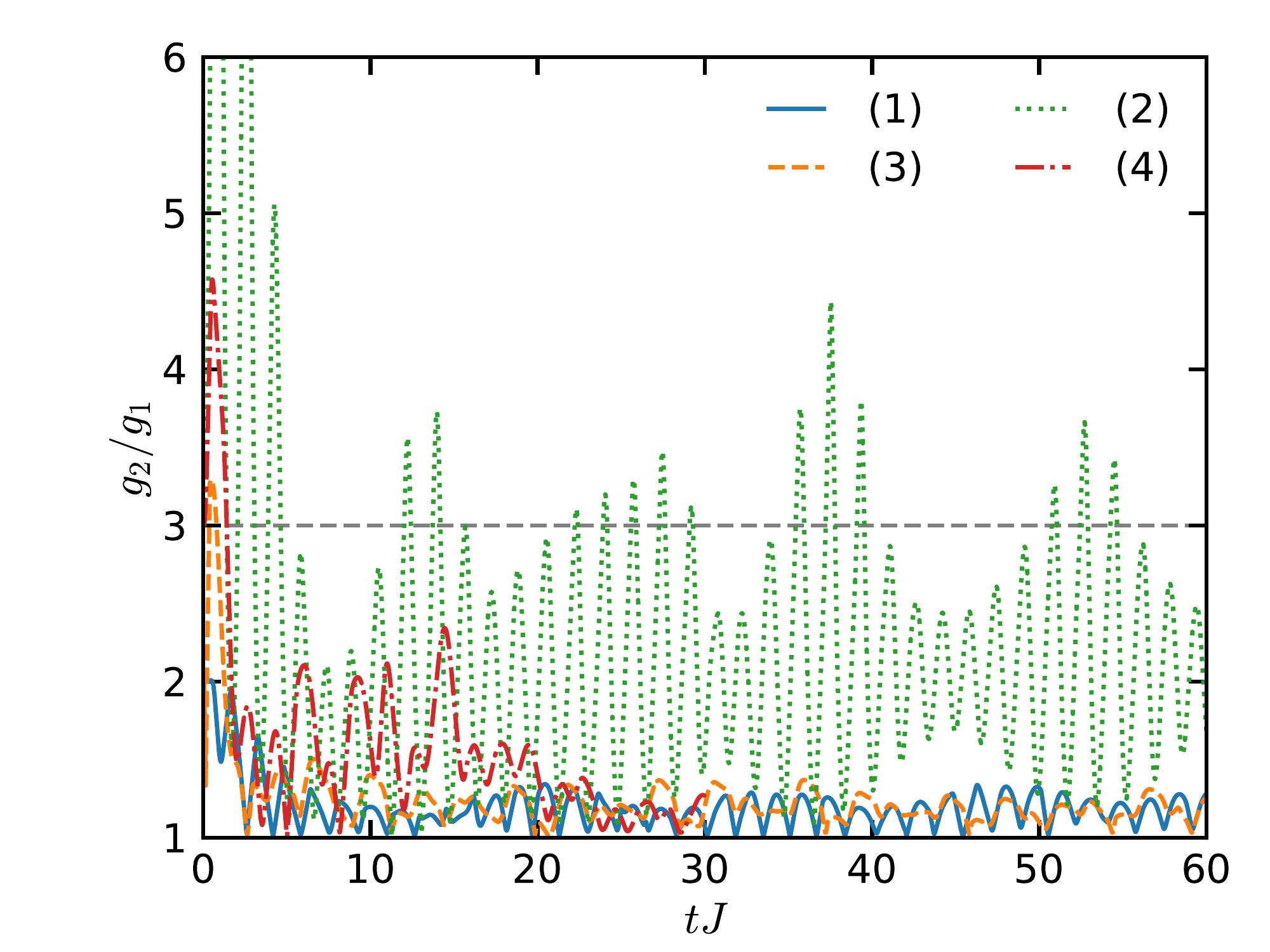}
    \caption{Entanglement gap ratio $g_2/g_1$ for all mesonic quench types (1-4) of Fig.\,1. The gray dashed background line denotes the corresponding CFT value.}
    \label{fig:ratios_comparison}
\end{figure}

In Fig.\,\ref{fig:ratios_comparison}, the entanglement gap ratios $g_2/g_1$ are compared for all mesonic quench types. The values are decaying at early times after the initial quench phase and oscillate around some mean value below the corresponding CFT value, which is shown as a reference scale by the gray dashed line. The oscillations are broken in all cases by cusps at the lower bound $g_2/g_1=1$. Similar nonanalyticities exist also in higher order ratios and eigenvalues of the modular Hamiltonian. For that reason, we concluded in the main text that the meson content of the QMB system or QFT is solely and fully contained in the dominant eigenvalue $\xi_0$, which gives rise to the phenomenon of entanglement oscillations.

\section{Prony signal analysis method}
\label{app:Prony}

Prony methods \cite{peter2014generalized} are based on the representation of a function $f(t)$ as a sum of complex exponentials, $f(t) = \sum_{k=1}^K c_k \e^{-i \, \omega_k\, t}$. The complex coefficients $c_k$ and frequencies $\omega_k$ can be determined independently through the linearity of the ansatz ($K$ is the (variable) total number of modes). This allows to capture both oscillations (real part of $\omega_k$) and exponential decay/growth (imaginary part of $\omega_k$). Here, we employ a signal analysis technique developed in \cite{Banuls:2019qrq}, for which the Prony method is applied on a finite analysis window that is sequentially shifted towards later times. Identified modes are visualized as poles in the complex frequency plane (on a color scale from blue to red). Stable modes can be interpreted as discrete frequencies, while streaks (a sequence of poles in different time windows) can be associated with branch cuts.


\end{document}